\documentclass[a4paper,12pt]{article}
\usepackage{amsmath,amssymb,amsthm,mathrsfs,graphicx,float,colordvi,url,wrapfig,color} 
\usepackage{enumerate,enumitem,here,mathrsfs,wrapfig,ulem,comment}   

\def\nn{\nonumber \\}  
 
\def\bg{\boldsymbol{g}}
\def\bga{\boldsymbol{\gamma}}
\def\bF{\boldsymbol{F}}   
\def\bp{\boldsymbol{\phi}} 
\def\bomg{\boldsymbol{\omega}} 
\def\bx{\bar{x}}
\def\dl{\partial}
\def\dlx{\partial_x}

\begin{document} 
\renewcommand{\thefootnote}{\fnsymbol{footnote}} 
\begin{titlepage}

\vspace*{0.0mm}

\begin{center}
{\large \bf 
Coarse-graining of the Einstein-Hilbert Action}
\\
\vspace*{2.5mm} 
{\large \bf 
rewritten by the Fisher information metric}
\\
\vspace*{9mm} 

\normalsize
{\large {Shingo Takeuchi\footnote{shingo.portable(at)gmail.com}}} 

\vspace*{4mm} 

\textit{
$^2$Institute of Research and Development, Duy Tan University}\\
\vspace*{0.5 mm}
\textit{P809, 3 Quang Trung, Hai Chau, Da Nang, Vietnam}\\
\vspace*{2.0 mm}
 
\end{center}

%-------------------------------
\vspace*{1mm}
\begin{abstract}
In this study, considering the Fisher information metric (Fisher metric) given by a specific form, 
which is the form of weights in statistical physics, we rewrite the Einstein-Hilbert (EH) action. 
Then, determining the transformation rules of the Fisher metric, etc under the coarse-graining, 
we perform the coarse-graining toward that rewritten EH action. 
We finally show an existence of a trivial fixed-point. 
Here, the existence of a trivial fixed-point is not trivial for us because we consider the metric 
given by the Fisher metric, which is not the normal metric and has to satisfy some constraint 
in the formalism of the the Fisher metric. We use the path-integral in our analysis. At this time 
we have to accept that a fundamental constraint in the formalism of the Fisher metric is broken 
at the quantum level. However we consider we can accept this with the thought that some constraints 
and causal relations held at the classical level usually get broken at the quantum level. We finds some 
problems of the Fisher metric. The space-time we consider  is two-dimensional. 
\end{abstract}

\end{titlepage}

%================================================================================================================== 
\section{Introduction}   
\label{Chap:Intro}
%================================================================================================================== 

Behavior of the space-time at the scale where the quantum effect becomes dominant is very important for us. 
To understand it, the quantum theory of gravity is essential. One of the interesting momenta 
for this direction are to reconsider the thermodynamics 
\cite{Fulling:1972md, Davies:1974th, Unruh:1976db,Hawking:1974sw} and holographic nature 
\cite{tHooft:1993dmi, Susskind:1994vu, Bigatti:1999dp, Bousso:2002ju} originally inhered in the gravity. We here would like 
to focus on the two works in this direction:
\begin{enumerate}
\renewcommand{\labelenumi}{\arabic{enumi}).}
\item 
One is \cite{Jacobson:1995ab}, which illustrates the derivation of the Einstein equation from the first law of thermodynamics 
by considering the local Rindler horizon at each point of a space-time.  
\item 
Another one is \cite{Verlinde:2010hp}, which proposes that the space-times are composed of some layers. 
Each layer is analog of the event-horizon, so the area of which is proportional to the amount of entropy associated with the inside of that layer.
This idea sheds light on the entropy to be the origin of the gravitational force. 
\end{enumerate}
Thermodynamics and entropy play the role of the fundamental ground in the above studies. These are the result of the statistical average,  
and looking above studies we would be led to a notion that the gravitational theory might be an effective theory given from some statistical averages. 
From this standpoint, it is one of very interesting attempts to consider the description of  the gravitational theory by the Fisher information 
metric (Fisher metric).

Based on the notion above, in the former part in this paper, we will rewrite the Einstein-Hilbert (EH) action with the Fisher information 
metric. This is because the Fisher metric could go along with the notion above since the metrices are given as a statistical average 
in the theory of the Fisher metric.  Actually, it would be an interesting attempt to check how it will be if explicitly rewriting EH action in terms 
of Fisher metric.

Here, the form of the Fisher metric we consider is not the original general one (given in (\ref{DefFishmet1})) 
but a specific one based on a the statistical physics (concretely see Sec.\ref{Sec:EPG} and \ref{Sec:FM}).   
We consider that assuming such a specific form would be allowed based on the two fact: 
1) the Fisher metric is originally given from a statistical average as we mention under (\ref{dforgp}), and 
2) a fact that it is a convenient form for the feasibility of analysis at present.

After rewriting the EH action in terms of the Fisher metric, we determine the transformation rules of the ingredients 
in the rewritten EH action under the coarse-graining, and perform the coarse-graining toward our rewritten 
EH action to investigate the fixed-point.

What we will do in this study are finally the following three:  
\begin{enumerate}
\renewcommand{\labelenumi}{\arabic{enumi}).}
\item rewrite the EH action in terms of the ingredients in the two-dimensional Fisher metric, 
\item determining the transformation rules of the Fisher metric, etc under the coarse-graining, perform the coarse-graining toward that rewritten EH action,  
\item based on that, obtain a trivial fixed-point and Newton coupling constant at that trivial fixed-point. 
\end{enumerate}

Here, the existence of a trivial fixed-point is not trivial for us because we consider 
the metric given by the Fisher metric, which is not the normal metric and has to satisfy 
some constraint in the formalism of the the Fisher metric.

In our analysis we consider the path-integral to involve the quantum effect, 
which leads to a critical problem that one of fundamental relations in the formalism of 
the Fisher metric is broken  (for more detail, see the end of Sec.\ref{SSec:RLS}). 
However, it would be also true that we should involve quantum effect in order to consider 
the realistic situation, thus we would have to accept such a critical situation that one of 
fundamental relations in the formalism of the Fisher metric is broken at the quantum level 
in the gravitational theory of the Fisher metric. 
At this time, we consider that some constraints and causal relations held at the classical 
level usually get broken at the quantum level for the quantum uncertainty.  
Anyway, this makes us notice that it is not a straightforward 
to involve the quantum effect in the gravitational theory of the Fisher metric. 

We also notice a problem that the degree of freedom of the Fisher metric cannot satisfy the degree of freedom that the metric in the general relativity has  (see the end of Sec.\ref{Sec:What} for more detail).   
\newline

Here, the existence of a trivial fixed-point is not trivial for us because we consider the metric 
given by the Fisher metric, which is never the normal metric and has to satisfy some constraint 
in the formalism of the the Fisher metric.

Renomalizations and fixed-points (and asymptotic safeties (fixed-points at ultra-violet fixed-points)) 
in gravitational theories have been studied until now to unveil how the gravitational theories will be 
and the quantum corrections of those will be at the ultraviolet region.  
%---

As for the EH actions, we would like readers to refer review papers 
\cite{Niedermaier:2006wt,Niedermaier:2006ns,Percacci:2007sz,Reuter:2012id,Codello:2008vh,Eichhorn:2017egq,Pereira:2019dbn} 
since number of papers for that is so huge.  
%---
the EH actions in various modified gravitational theories have been also progressed. 
We would like to refer some of those in the range author can refer. 

As for the EH action with the higher order derivative corrections, 
\cite{Ohta:2011rv,Muneyuki:2012ur,Ohta:2012vb,Ohta:2013uca} and 
\cite{Narain:2012nf,Narain:2012te,Narain:2013eea}.
As for the $f(R)$ gravity, \cite{Ohta:2015efa,Ohta:2015fcu,Falls:2016msz,Ohta:2018sze,deBrito:2018jxt}. 
Also \cite{Falls:2010he,Falls:2013bv,Falls:2014zba,Falls:2014tra,Falls:2015qga,Falls:2015cta,Falls:2016wsa,Falls:2016msz,Falls:2017cze,Falls:2017lst} 
for black hole backgrounds,  $f(R)$ gravities, Newton constants, cosmological constants and so on in the asymptotic safety.

As for some modified gravities, \cite{Muneyuki:2013aba,Ohta:2015zwa,Ohta:2020bsc}, 
\cite{Narain:2009fy,Narain:2009gb,Park:2019lkh} and \cite{Park:2014tia,Park:2017wiw}. 
As for phenomenological models coupling with gravities, 
\cite{Oda:2015sma,Wetterich:2016uxm,Hamada:2017rvn,Eichhorn:2017als,Pawlowski:2018ixd,deBrito:2019epw,Wetterich:2019zdo}.

We mention some relation between above other studies and this study. 
Since the model we treat in this study is just a EH action, 
things written in the review papers \cite{Niedermaier:2006wt,Niedermaier:2006ns,Percacci:2007sz,Reuter:2012id,Codello:2008vh,Eichhorn:2017egq,Pereira:2019dbn} 
are related with this study. 
However, this study finally becomes two-dimensional, and there is no description about the results concerning two-dimensional in these. 
However, central issues in the business of renomalizations are whether unitarity and renomalizability are respectively held or not and how 
fixed-points are. As for the unitarity and renomalizability, it can be considered 
that both would be held in our study, since the model we consider in this study is without higher order derivative corrections (which generally break the unitarity) 
and is lower dimensional.  On the other hand, we could not compare the fixed-point we will get in this study with that in the results in the above 
review paper. This is because our coarse-graining will proceed to lower energy direction 
(generally speaking, coarse-graining is to look the low energy and long range behavior  
of the system by summarizing up the fine behaviors)  
and the fixed-point we get in this study is the one at the low energy limit, which is opposite from those in the review papers. 
 
There is a series of papers for the study on the $D=(2+\epsilon)$ EH action 
exploiting the renormalization theory \cite{Kawai:1989yh, Kawai:1992np, Kawai:1993mb, 
Aida:1994np,Aida:1995ah,Aida:1994zc, Kitazawa:1995ru,Kitazawa:1995gc,Kawai:1995ju,Aida:1996zn}. 
%--- 
Our and their physical results are hoped to agree each other as long as physically the same D=2 gravity by EH action  
(though how they and we treat the gravity are fundamentally different; they are by the metric-base and we are 
by rewriting the metric by the Fisher metric). However, to check it and make a conclusion on this consistency is difficult. 
This is because they study the high energy region  (and technical problems) of the $D=2$ EH action, on the other hand we 
study the low energy region of that as we perform the coarse-graining, therefore their and our results are in physically different 
energy region. 

However, even if our and their results came out differently each other, it may not be problems. 
This is because they perform their studies based on the metric, whereas we perform our study 
by rewriting the metric by the Fisher metric,  which brings the following 2 points into the analysis: 
1) the constraint the Fisher metric should satisfy, which is (\ref{sump1}) (moreover it is broken at the quantum level as mentioned in Sec.\ref{SSec:RLS}),  
2) the problem for the d.o.f. as we write in the end of Sec.\ref{Sec:What}. 
These 2 points may change the fundamental properties in the gravitational theory, 
and affects the final physical results. (There is also one more problem in our study, which is that  
we suppose the form of the Fisher metric as in (\ref{dforgp}), but generally various forms other than 
it can be considered in principle.)  If our and their results came out differently, we would consider that 
our study is a report for the EH action of the Fisher metric.  

Lastly, 2D quantum gravity given by the Polyakov action is studied in \cite{Polyakov:1987zb,Knizhnik:1988ak,Distler:1988jt}. 
In what follows, we mention the organization of this paper.  
\newline

In Sec.\ref{Sec:What}, we define the Fisher metric with the probability distribution $p(x,\theta) = e^{-\gamma(x,\theta)}$ 
($x$ and $\theta$ mean labels of states and parameters). We also touch on the point that   
the number of the degree of freedom of the Fisher metric is not $D(1+D)/2$ but $D$.

In Sec.\ref{Sec:EPG}, we specify the form of $p$ as $p(x,\theta)=e^{-\left(\theta^\mu \bF_\mu (x) - \phi(\theta)\right)}$,   
and then give the expression of the Ricci tensor in terms of that $p$. 

In Sec.\ref{Sec:FM}, specifying $\theta^\mu$, $\bF_\mu$ and $\phi$ in the $p$ given in Sec.\ref{Sec:EPG}, 
we obtain the expression of the Ricci tensor and then the EH action described by those.     
 
In Sec.\ref{Sec:Renomalization}, explaining for the space in which we define our rewritten EH action 
and introducing the fixed-points, we give the transformation rules for the ingredients in the EH action under 
the coarse-graining. Particularly in Sec.\ref{SSec:RLS}, we mention the crucial problem in the path-integral 
for the Fisher metric. 
 
In Sec.\ref{Sec:EHAofrt}, we perform the coarse-graining toward our rewritten EH action. 
From that, we obtain a trivial fixed-point and Newton coupling constant at that fixed-point in our rewritten EH action.

In Appendices.\ref{App:CalEin} and \ref{App:CalEinbga}, 
we derive the expression of the Ricci tensors in the case with $p=e^{-\gamma}$.   
Then based on that expression, we derive the expression of the Ricci tensors when $p$ is given as $p=e^{-\left(\theta^\mu \bF_\mu  - \phi \right)}$. 
The final result is (\ref{nzlsmt}), which leads to (\ref{Xmntr3}). 
\newline

Lastly, a paper \cite{Matsueda:2013saa} rewrites the Einstein equation using the Fisher metric. 
Since it is the same topic with the former part of this paper, 
we should touch on the differences between \cite{Matsueda:2013saa} and our paper. 
It is true that a large part of the analysis in \cite{Matsueda:2013saa} is very helpful for the analysis in our study,  
however it performs an ill-justified analysis, which is between (76) and (77) in \cite{Matsueda:2013saa}; 
It is a manipulation such as $\left< {\cal O}_1 {\cal O}_2 \right> = \left< {\cal O}_1 \right>\left< {\cal O}_2 \right>$ 
(${\cal O}_{1,2}$ mean some observables). 
Although \cite{Matsueda:2013saa} mentions it is approximation,  
what sense \cite{Matsueda:2013saa} is saying as approximation is unclear.   
Our analysis, specifically in Sec.\ref{Sec:FM}, 
is performed without that manipulation.

%==================================================================================================================  
\section{Definition of the Fisher metric and points to be careful}
\label{Sec:What} 
%================================================================================================================== 
 
In this section, we give the definition of the Fisher information metric (Fisher metric).    
Let us first consider a statistical theory with parameters $\theta \equiv (\theta^1, \theta^2, \cdots, \theta^D)$. 
Then let us consider probability distribution $p_x(\theta)$, 
where $x$ means the label distinguishing physical states. 
The summation of $p$ is generally written as $\sum_{x=0}^\infty p_x(\theta) =1$. 
We can change the labeling of $x$ without loss of generality as $\sum_{x=-\infty}^\infty p_x(\theta) =1$. 
Let us consider the case that $x$ is continues numbers. 
Then we can write this relation as
\begin{align}\label{sump1}
\int_{-\infty}^\infty \! dx \, p(x,\theta)=1.   
\end{align}
Note that now the $1$ above is a constant never depending on $\theta$.

We here would like to give the definition of the statistical average. 
If we write $\langle \cdots \rangle$, which means 
\begin{align}\label{ladddrlstav}
\langle \cdots \rangle 
\, \equiv \, 
\frac{\int_{-\infty}^\infty \! dx \,\cdots \, p(x,\theta)}{\int_{-\infty}^\infty \! dx \, p(x,\theta)} 
= \int_{-\infty}^\infty \! dx \,\cdots \, p(x,\theta),
\end{align}
where since $\int_{-\infty}^\infty \! dx \, p(x,\theta) = 1$ as in (\ref{sump1}), 
we do not need to write the denominator explicitly. 
In what follows, we basically abbreviate to write the integral region of $x$.

Let us represent $p$ as
\begin{align}\label{defpth}
p(x,\theta) = e^{-\bga(x,\theta)}. 
\end{align}
Then, the Fisher metric is defined as
\begin{align}\label{DefFishmet1}
g_{\mu\nu}(\theta)
=\int \! dx \, p \frac{\dl \bga(x,\theta)}{\dl \theta^\mu} \frac{\dl \bga(x,\theta)}{\dl \theta^\nu}
=\left< \dl_\mu \bga(x,\theta) \dl_\nu \bga(x,\theta) \right>, \quad \dl_\mu \equiv \frac{\dl}{\dl \theta^\mu}. 
\end{align}
where $\mu, \nu =1,2,\cdots, D$.  
Therefore, based on the Fisher metric, 
we can consider an $n$-dimensional Riemannian geometry with $\theta$ as the coordinates.

In what follows, we denote $\gamma$ and $g_{\mu\nu}$ before taking the statistical average in boldface. 
Therefore, $\bg_{\mu\nu} = \, \dl_\mu \bga \dl_\nu \bga$ and $g_{\mu\nu} = \langle \bg_{\mu\nu} \rangle$.

One way to obtain the squared line-element with the Fisher metric is the difference of the probability distribution 
for infinitesimal variations of $\theta$. We show it as
\begin{align}\label{infdis}
\int \! dx \, \Big(p(x,\theta+d\theta) - p(x,\theta) \Big)
=& \,\, g_{\mu\nu}(\theta) \, d\theta^\mu d\theta^\nu +{\cal O}(d\theta^3),
\end{align}

In addition to (\ref{DefFishmet1}), there is another expression for the Fisher metric as\footnote{
% ********** FOOTNOTE ********** 
To show (\ref{DefFishmet3}), let us start with
\begin{eqnarray}\label{octprosw} 
\dl_\mu (p \, \dl_\nu \bga) = \dl_\mu p \, \dl_\nu \bga + p \, \dl_\mu \dl_\nu \bga. 
\end{eqnarray}
Using $p \, \dl_\nu \bga = - \dl_\nu \, p$, we can rewrite (\ref{octprosw}) as
\begin{eqnarray}\label{octprorw}
-\dl_\mu \dl_\nu p = - p \, \dl_\mu \bga \dl_\nu \bga + p \, \dl_\mu \dl_\nu \bga. 
\end{eqnarray}
Since $\int \! dx \, p$ is unit, a constant, $\int \! dx \, \dl_\mu \dl_\nu \, p =0$ and we can reach (\ref{DefFishmet3}). 
Note that $\langle \dl_\mu \dl_\nu \bga \rangle = \langle \dl_\mu \bga \dl_\nu \bga \rangle$ 
but $\dl_\mu \dl_\nu \bga \not= \dl_\mu \bga \dl_\nu \bga$.
% ****************************** 
}  
\begin{align}\label{DefFishmet3} 
g_{\mu\nu} = \, \left< \dl_\mu \dl_\nu \bga \right>. 
\end{align}
 
Lastly, we would like to give attention to the problem for d.o.f. of Fisher metric.  
\newline

We can see from (\ref{DefFishmet1}) that the number of the degree of freedom of the fisher metric is $D$ 
($D$ means the number of components of $\theta^\mu$). 
Since (\ref{DefFishmet1}) is the general definition of the the fisher metric, 
we can see that the number of the degree of freedom of the Fisher metric 
is not $D(1+D)/2$ but $D$ in general, and is usable when it has no non-diagonal components. 
Since this problem is rooted in the definition of the Fisher metric, 
this would be an unavoidable problem as long as we consider the Fisher metric along its definition\footnote{
If we cure this problem, 
since this is the problem of the lack of the number of $\theta^\mu$, 
we have to add more $\theta^\mu$. For example, we consider the following way in the four-dimensional case:
\begin{align}
g_{\mu\nu}  = \left(
    \begin{array}{cccc}
\left< \dl_{00} \bga(x,\theta) \dl_{00} \bga(x,\theta) \right> & 
\left< \dl_{00} \bga(x,\theta) \dl_{01} \bga(x,\theta) \right> & 
\left< \dl_{00} \bga(x,\theta) \dl_{02} \bga(x,\theta) \right> & 
\left< \dl_{00} \bga(x,\theta) \dl_{03} \bga(x,\theta) \right> \\
- & 
\left< \dl_{11} \bga(x,\theta) \dl_{11} \bga(x,\theta) \right> & 
\left< \dl_{11} \bga(x,\theta) \dl_{12} \bga(x,\theta) \right> & 
\left< \dl_{11} \bga(x,\theta) \dl_{13} \bga(x,\theta) \right> \\
- & - & 
\left< \dl_{22} \bga(x,\theta) \dl_{22} \bga(x,\theta) \right> & 
\left< \dl_{22} \bga(x,\theta) \dl_{23} \bga(x,\theta) \right> \\
- & - & - & 
\left< \dl_{33} \bga(x,\theta) \dl_{33} \bga(x,\theta) \right> 
    \end{array}
  \right)
\end{align}
where 
$\theta^\mu=(
\theta^{00}, \theta^{01}, \theta^{02}, \theta^{03}, 
\theta^{11}, \theta^{12}, \theta^{13}, \theta^{22},  
\theta^{23}, \theta^{33}
)$ and $\dl_{\mu\nu} \equiv \frac{\dl}{\dl \theta^{\mu\nu}}$. 
}. 

There is another problem that (\ref{sump1}) gets broken when we consider the quantum effect of the Fisher metric, 
which we mention in Sec.\ref{SSec:RLS}.

%==================================================================================================================  
\section{Form of $p$ and the Ricci tensor in terms of that}
\label{Sec:EPG} 
%================================================================================================================== 

In the previous section, we have given the definition of the Fisher metric given 
by the statistical average with the probability distributions $p=e^{-\bga}$ 
as in (\ref{DefFishmet1}) and (\ref{DefFishmet3}).  
In this section, we specify the form of $\bga$ as in (\ref{dforggm}), 
and then give the expression of the Ricci tensor in that $p$ as in (\ref{Xmntr3}).  
\newline

From the general statistical physics's point of view, we will write $p(x,\theta)$ as $p(x,\theta)=e^{ -\beta E(x,\theta) - \ln Z(\theta)}$. 
Then as the generalization of this form, we can consider various forms for $p(x,\theta)$. 
Among these, we consider the most simple form but keeping essential features as the statistical physics as 
\begin{align}\label{dforgp}
p(x,\theta)=e^{ \theta^\mu \bF_\mu (x) - \phi(\theta)}. 
\end{align}
Considering the original physical rule of $p(x,\theta)$ given in the previous subsection,  
expressing $p(x,\theta)$ in the way of (\ref{dforgp}) would be allowed as one of possible 
forms. As for other forms possible to consider, for example we can take 
$\gamma(x,\theta)
=\theta^\mu \theta^\nu G_{\mu\nu}(x)
-\theta^\mu F_\mu (x)- \phi(\theta)$.

If  $p(x,\theta)$ is given as (\ref{dforgp}), 
$\bga$ in (\ref{defpth}) can be written as 
\begin{align}\label{dforggm} 
\bga(x,\theta) = - \theta^\mu \bF_\mu (x) + \phi(\theta).
\end{align}  
At this time, the differentials of $\bga$ are given as 
\begin{eqnarray}
\label{ggdrl1}
\dl_\mu \bga(x,\theta) \!\! &=& \!\! -\bF_\mu (x) + \dl_\mu \phi(\theta), \\
[1.5mm]
\label{ggdrl2}
\dl_\mu \dl_\nu \bga(x,\theta) \!\! &=& \!\! \dl_\mu \dl_\nu \phi(\theta).
\end{eqnarray} 
Note that when $p$ is assumed as in (\ref{dforgp}), as can be seen in (\ref{ggdrl2}),  
$\bg_{\mu\nu}(x,\theta)$ is independent of $x$ in the stage before the statistical average is taken.
Therefore, in the case of (\ref{dforgp}), the following manipulation is possible: 
\begin{align}\label{ssddgm} 
\left< \dl_\mu \dl_\nu \bga(x,\theta) \, \cdots \right> 
= \dl_\mu\dl_\nu \phi(\theta) \left< \cdots \right>  
= g_{\mu\nu}(\theta) \left< \cdots \right>.
\end{align}  
Therefore, when $p$ is assumed as (\ref{dforgp}), we can express $g_{\mu\nu}(\theta)$ without $\langle \cdots \rangle$ as 
\begin{align}\label{metexpfm}
\left< \dl_\mu \dl_\nu \bga(x,\theta) \right> 
= \dl_\mu\dl_\nu \phi(\theta) 
= g_{\mu\nu}(\theta).
\end{align}

From (\ref{ggdrl1}), the statistical average of $\bF_\mu(x)$ can satisfy the following relation: 
\begin{align}\label{Nmpnp} 
\left< \bF_\mu(x) \right> = \dl_\mu \phi(\theta), 
\end{align}
where we have used $\int \! dx \, \dl_\mu \bga \, p = - \dl_\mu \left( \int dx \,p \right)= 0$. 
Then, from (\ref{ggdrl1}), we can see 
\begin{eqnarray}
\left< \dl_\mu \bga\dl_\nu \bga \right> 
\!\! &=& \!\! 
\left< \bF_\mu \bF_\nu \right>
- \left<\bF_\mu\right> \dl_\nu \phi 
- \left<\bF_\nu \right> \dl_\mu \phi 
+ \dl_\mu \phi \, \dl_\nu \phi, 
\nn
[1.5mm]
\!\! &=& \!\! \left<\bF_\mu \bF_\nu\right> - \left<\bF_\mu\right>\left<\bF_\nu\right>,
\end{eqnarray}
where we have used (\ref{Nmpnp}). 
Therefore, when $p$ is given as in (\ref{dforgp}), 
in addition to (\ref{DefFishmet1}) and (\ref{DefFishmet3}), 
we have another expression of the Fisher metrices as
\begin{align} \label{exsgppFF}
g_{\mu\nu}
=   \, \left<\bF_\mu \bF_\nu \right> - \left<\bF_\mu\right> \left<\bF_\nu \right>
= - \left< \bF_\mu \dl_\nu \bga \right>.
\end{align}
\newline

In (\ref{ExpRicibga}), we have given the expression of the Ricci tensor in $p=e^{-\bga}$.  
Through the derivation we note in Appendix.\ref{App:CalEinbga}, 
we can obtain the expression of the Ricci tensor in $\bga=\theta^\mu \bF_\mu - \phi$ as 
\begin{eqnarray}\label{Xmntr3}
R_{\mu\nu} 
\!\! &=& \!\! \hspace{2mm}
\frac{1}{4}g^{\sigma\tau}g^{\rho\zeta}
\left( 
\left< \bF_{\mu}\dl_\zeta \bga \dl_\sigma \bga \right> \left< \bF_{\nu} \dl_\rho \bga \dl_\tau \bga \right> 
- \left<\bF_{\mu}\right>g_{\zeta\sigma} \left<\bF_{\nu}\right>g_{\rho\tau}
\right) \nonumber \\ 
[1.5mm]
&& \!\!\!\!
- \frac{1}{4}g^{\sigma\tau}g^{\rho\zeta}
\left< \dl_{\rho}\bga\dl_{\sigma}\bga\dl_{\tau}\bga \right> \left< \bF_{\mu} \bF_{\nu}\dl_{\zeta}\bga \right>. 
\end{eqnarray}

%==================================================================================================================  
\section{Specifying of $\theta^\mu$, $\bF_\mu$ and $\phi$, and rewriting of the EH action}
\label{Sec:FM} 
%================================================================================================================== 

In the previous section, we have obtained the expression of the Ricci tensor when $p=e^{-\bga}$ with $\bga=-\theta^\mu \bF_\mu + \phi$, 
as in (\ref{Xmntr3}). In this section, specifying $\theta^\mu$, $\bF_\mu$ and $\phi$, 
we obtain the expression of the Ricci tensor and then the Einstein-Hilbert (EH) action described by those.      

%==================================================================================================================  
\subsection{$\theta^\mu$, $\bF_\mu$ and $\phi$} 
\label{Sec:EPO} 
%==================================================================================================================  

We consider the following $\theta^\mu$, $\bF_\mu$ and $\phi$: 
\begin{eqnarray}
\label{pAdS2FmFbt}  
\bF(x) \!\! &=& \!\! (\bF_1(x), \bF_2(x))=(x,x^2),\\
[1.5mm]   
\label{pAdS2FmTbt}
\theta \!\! &=& \!\! (\theta^1, \theta^2) = \left(  \frac{\bx}{\sigma_0^2}, - \frac{1}{2\sigma_0^2} \right),\\ 
[1.5mm]
\label{pAdS2FmPbt} 
\phi(\theta) 
\!\! &=& \!\! \ln \left[\sqrt{2\pi c_1} \sigma_0 \right] + \frac{\bx^2}{2c_2\sigma_0^2} 
= \frac{1}{2} \ln \left[ -\frac{\pi}{\theta^2} \right] - \frac{1}{4c_2}\frac{(\theta^1)^2}{\theta^2},  
\end{eqnarray}
where the dimension of $\sigma_0^2$ is supposed to be the dimension of length$^{-1}$
and $c_{1,2}$ are both some quantities with the dimension of length so that $\theta^{1,2}$ can have the dimension of length.  
We put  $c_{1,2}$ to $1$ for the simplicity of analysis in what follows.  
Then, $p$ in (\ref{dforgp}) can be written as
\begin{align}\label{pAdS2Fmbt} 
p
= \exp \left[ - \ln \left[\sqrt{2\pi}\,\sigma_0 \right] - \frac{x^2}{2\sigma_0^2} 
+ \frac{x\bx}{\sigma_0^2} - \frac{\bx^2}{2\sigma_0^2} \right]
= \frac{1}{\sqrt{2\pi}\,\sigma_0} \, \exp \left[ -\frac{(x-\bx)^2}{2\sigma_0^2} \right].   
\end{align} 
We here would like to comment that $\bx$ and $\sigma_0$ can be taken freely, therefore 
$\theta^\mu$ can form a $2$-dimensional Riemannian geometry with the coordinates $\theta^\mu$. 
This is because the number of the free parameters to determine $\theta^\mu$ and the number of 
$\theta^\mu$ can agree each other. Here, note that since $p$ is given like (\ref{pAdS2Fmbt}), 
no matter how $\bx$ and $\sigma_0$ are taken, (\ref{sump1}) is always kept.

When $p$ is given as in (\ref{pAdS2Fmbt}), the statistical system behind the Fisher metric have the following two properties: 
1) The realizing state as a result of the statistical average is a state labeled by a $\bx$.  
2) The frequency of the appearance of the states labeled by $x$ follows the Gaussian distribution around $\bx$. 
\newline

When $p$ is given as in (\ref{pAdS2Fmbt}), the Fisher metric is composed as 
\begin{align}\label{mt22x00}
\left( 
\begin{array}{cc}
g_{00} & g_{01} \\ - & g_{11} 
\end{array}
\right) = \frac{1}{2} \,
\left(
\begin{array}{cc}
\displaystyle -\frac{1}{\theta^2} & \displaystyle \frac{\theta^1}{(\theta^2)^2} \\
\displaystyle -                   & \displaystyle \frac{-(\theta^1)^2+\theta^2}{(\theta^2)^3}
\end{array}
\right), \quad
%----
\left( 
\begin{array}{cc}
g^{00} & g^{01} \\ - & g^{11} 
\end{array}
\right) = 2 \,
\left(
\begin{array}{cc}
\displaystyle (\theta^1)^2-\theta^2 & \displaystyle \theta^1 \theta^2 \\
\displaystyle -                     & \displaystyle (\theta^2)^2
\end{array}
\right) 
\end{align}
We can reach the one above if we calculate based on (\ref{DefFishmet1}) or (\ref{metexpfm}). 
Lastly we shall give our attention to the point given at the last in Sec.\ref{Sec:What}.

%================================================================================================================== 
\subsection{EH action}
\label{Sec:EHAction} 
%================================================================================================================== 

In this subsection, we obtain the EH action described by  $\theta^\mu$, $\bF_\mu$ and $\phi$.  
We first rewrite a part, $\left< \bF_\mu \dl_\zeta \bga \dl_\xi \bga \right>$, in (\ref{Xmntr3}). 
To this purpose, we expand as
\begin{eqnarray}
%----------  
\label{bFmuTyl}
\bullet && \!\!\! \!\!\!
\bF_\mu(x) = \bF_\mu(\bx)+\dlx \bF_\mu(x)\big|_{x=\bx}(x-\bx) + \frac{1}{2}\dlx^2 \bF_\mu(x)\big|_{x=\bx}(x-\bx)^2, 
\\
%----------
[1.5mm]
\label{gGmuTyl}  
\bullet && \!\!\! \!\!\!  
\dl_\mu \bga(x,\theta) = \dl_\mu \bga(x,\theta)\big|_{x=\bx} + \dlx\dl_\mu \bga(x,\theta)\big|_{x=\bx}(x-\bx)
+ \frac{1}{2} \dlx^2\dl_\mu\bga(x,\theta)\big|_{x=\bx}(x-\bx)^2.\nn
%----------
\end{eqnarray} 
Note that there is no notation, ${\cal O}\left((x-\bx)^3\right)$, in the ones above. 
We can understand this from the fact that more than 3rd-order derivative with respect to $x$ vanish as
\begin{eqnarray} \label{dlnxdlmbgvz} 
\dlx^n \dl_\mu \bga(x,\theta)\big|_{x=\bx} 
&=&
-\dlx^n \bF_\mu (x)\big|_{x=\bx} \,=\, 0 \quad \textrm{for $n=3,4,5,\cdots$,} 
\end{eqnarray} 
when $\bga$ and $\bF_\mu$ are given as in (\ref{dforggm}) and (\ref{pAdS2FmFbt}), respectively.

Substituting (\ref{gGmuTyl}) into the condition:
\begin{align}\label{FunZer} 
\left< \dl_\mu \bga \right> = - \dl_\mu \left(\int \! dx \, p(x,\theta)\right)= 0.  
\end{align} 
we can obtain the following relation:
\begin{eqnarray}\label{exgFmuTyl}  
\frac{1}{2} \, \dlx^2\dl_\mu\bga(x,\theta)\big|_{x=\bx} \, \sigma_0^2 
\!\! &=& \!\! -\dl_\mu \bga(x,\theta)\big|_{x=\bx}.  
\end{eqnarray}
In the process to obtain the one above, we have used the following calculation:  
\begin{align}
%----------
\label{repx2bxtsgm2} 
\left< (x-\bx)^n \right> 
\, =& \,
\left\{
\begin{array}{cl}
0                     & \,\, \textrm{for $n=$ odd numbers,}  \\
[1.5mm]
(n-1)!! \, \sigma_0^2 & \,\, \textrm{for $n=$ even numbers,} 
\end{array}
\right.
\end{align}
where $\left< (x-\bx)^n \right> = \int dx\, (x-\bx)^n \, p$. 
The l.h.s. in (\ref{exgFmuTyl}) is the coefficient of the third term in the r.h.s. of (\ref{gGmuTyl}) (if divided by $\sigma_0^2$), 
In addition, we can also rewrite the coefficient of the second term in the r.h.s. of (\ref{gGmuTyl}) as
\begin{eqnarray}\label{exgFmuTng}  
\dlx \dl_\mu \bga(x,\theta)\big|_{x=\bx} 
= \dlx \dl_\mu \big( - \theta^\nu \bF_\nu (x) + \phi(\theta) \big) 
= - \dlx \bF_\mu (x). 
\end{eqnarray}

Using (\ref{bFmuTyl}) and (\ref{gGmuTyl}) with (\ref{exgFmuTyl}) and (\ref{exgFmuTng}), it turns out as
\begin{eqnarray}\label{FGGvan1}
\left< \bF_\mu(x) \dl_\zeta \bga(x) \dl_\xi \bga(x) \right> 
\!\! &=& \!\!
\hspace{0.0mm}
\Bigg< 
\left(
\bF_\mu(\bx) + \dlx \bF_\mu(x)\big|_{x=\bx}(x-\bx) + \frac{1}{2}\dlx^2 \bF_\mu(x)\big|_{x=\bx}(x-\bx)^2 
\right) \nn  
%----------
[1.5mm]
&&\hspace{-12.0mm} \times
\left\{
\dl_\zeta \bga(x,\theta)\big|_{x=\bx}\left( 1 - \frac{(x-\bx)^2}{\sigma_0^2} \right)  
-\dlx \bF_\zeta(x)\big|_{x=\bx}(x-\bx)
\right\} \nn 
%---------- 
%
[1.5mm]
&&\hspace{-12.0mm} \times
\left\{
\dl_\xi \bga(x,\theta)\big|_{x=\bx}\left( 1 - \frac{(x-\bx)^2}{\sigma_0^2} \right) - \dlx \bF_\xi(x)\big|_{x=\bx}(x-\bx)  
\right\}
\Bigg>. 
\end{eqnarray} 
The one above can be calculated as 
\begin{eqnarray} \label{FGGvank2}
(\ref{FGGvan1})
\!\! &=& \!\!
\left(2 \bF_\mu(\bx)+ 5 \sigma_0^2 \dlx^2 \bF_\mu(x)\big|_{x=\bx}\right) 
\dl_\zeta \bga(\bx,\theta) \dl_\xi \bga(\bx,\theta) \nn
%----------
[1.5mm]
&& \!\!\!\!
+ 2 \sigma_0^2 \dlx \bF_\mu(x)\big|_{x=\bx}
\Big(
\dlx \bF_\xi(x)\big|_{x=\bx} \dl_\zeta \bga(\bx,\theta) + \dlx \bF_\zeta(x)\big|_{x=\bx} \dl_\xi \bga(\bx,\theta)
\Big) \nn
%----------
[1.5mm]
&& \!\!\!\!
+
\dlx \bF_\zeta(x)\big|_{x=\bx}\dlx \bF_\xi(x)\big|_{x=\bx}
\left( \sigma_0^2 \bF_\mu(\bx) + \frac{3}{2}\sigma_0^4\dlx^2 \bF_\mu(x)\big|_{x=\bx} \right) \nn
%====================
[1.5mm]
\!\! &=& \!\!
\left( 2 \bF_\mu(\bx)+ 5 \sigma_0^2 \dlx^2 \bF_\mu(x)\big|_{x=\bx} \right) 
\Big(
- \bF_\zeta(\bx) \dl_\xi \phi(\theta) - \bF_\xi(\bx) \dl_\zeta \phi(\theta) + \dl_\zeta \phi(\theta) \dl_\xi \phi(\theta)
\Big) \nn
%----------
[1.5mm]
&& \!\!\!\!
+2 \sigma_0^2 \dlx \bF_\mu(x)\big|_{x=\bx}
\Big( \dlx \bF_\xi(x)\big|_{x=\bx} \dl_\zeta \phi(\theta) + \dlx \bF_\zeta(x)\big|_{x=\bx} \dl_\xi \phi(\theta) \Big)
+{\cal F}_{0,\mu\zeta\xi}, 
\end{eqnarray} 
where we have performed the calculations like 
$\left< \left(x-\bar{x}\right)^p \left(1-\frac{\left(x-\bar{x}\right)^2}{2\sigma_0^2}\right)^q \right>$,  
and when moving from the first to the second lines, we have used (\ref{dforggm}).  
Further, we have defined the $\theta$-independent constant part as
\begin{align}\label{thinptc}
{\cal F}_{0,\mu\zeta\xi}  
\equiv 
& \,
\left( 2 \bF_\mu(\bx)+ 5 \sigma_0^2 \dlx^2 \bF_\mu(x)\big|_{x=\bx} \right) \bF_\zeta(\bx) \bF_\xi(\bx) \nn
%----------
[1.5mm]
& \,
-2 \sigma_0^2 \dlx \bF_\mu(x)\big|_{x=\bx}
\Big( \dlx \bF_\xi(x)\big|_{x=\bx} \bF_\zeta(\bx) + \dlx \bF_\zeta(x)\big|_{x=\bx} \bF_\xi(\bx) \Big) \nn
%----------
[1.5mm]
& \,
+
\dlx \bF_\zeta(x)\big|_{x=\bx}\dlx \bF_\xi(x)\big|_{x=\bx}
\left( \sigma_0^2 \bF_\mu(\bx) + \frac{3}{2}\sigma_0^4\dlx^2 \bF_\mu(x)\big|_{x=\bx} \right).
\end{align}
We can see ${\cal F}_{0,\mu\zeta\xi} = {\cal F}_{0,\mu\xi\zeta}$.
Factorizing by $\dl_\xi \phi(\theta)$, $\dl_\zeta \phi(\theta)$ and $\dl_\zeta \phi(\theta) \dl_\xi \phi(\theta)$,
\begin{align}
\textrm{(\ref{FGGvank2})}
=& \,
\Big\{
- \left( 2 \bF_\mu(\bx)+ 5 \sigma_0^2 \dlx^2 \bF_\mu(x)\big|_{x=\bx} \right) \bF_\zeta(\bx)
+ 2 \sigma_0^2 \dlx \bF_\mu(x)\big|_{x=\bx} \dlx \bF_\zeta(x)\big|_{x=\bx}
\Big\}
\dl_\xi \phi(\theta) \nn
%----------
[1.5mm]
+& \,
\Big\{
- \left( 2 \bF_\mu(\bx)+ 5 \sigma_0^2 \dlx^2 \bF_\mu(x)\big|_{x=\bx} \right) \bF_\xi(\bx)
+ 2 \sigma_0^2 \dlx \bF_\mu(x)\big|_{x=\bx} \dlx \bF_\xi(x)\big|_{x=\bx}
\Big\}
\dl_\zeta \phi(\theta) \nn
%----------
[1.5mm]
+& \, 
\left( 2 \bF_\mu(\bx)+ 5 \sigma_0^2 \dlx^2 \bF_\mu(x)\big|_{x=\bx} \right) \dl_\zeta \phi(\theta) \dl_\xi \phi(\theta)
+{\cal F}_{0,\mu\zeta\xi} \nn
%----------
[1.5mm]
\equiv& \,\, 
{\cal P}_{\mu\zeta} \dl_\xi   \phi(\theta)
+{\cal P}_{\mu\xi}  \dl_\zeta \phi(\theta)
+{\cal Q}_\mu       \dl_\zeta \phi(\theta) \dl_\xi \phi(\theta)
+{\cal F}_{0,\mu\zeta\xi},
\end{align}
where ${\cal P}_{\mu\zeta}$ and ${\cal Q}_\mu $ are the $\theta$-independent constant parts given as 
\begin{eqnarray}
\label{dfPmr}
{\cal P}_{\mu\zeta} \!\! &\equiv& \!\!
- \left( 2 \bF_\mu(\bx)+ 5 \sigma_0^2 \dlx^2 \bF_\mu(x)\big|_{x=\bx} \right) \bF_\zeta(\bx) 
+ 2 \sigma_0^2 \dlx \bF_\mu(x)\big|_{x=\bx} \dlx \bF_\zeta(x)\big|_{x=\bx},\\ 
%-----
[1.5mm]
\label{dfQmr}
{\cal Q}_\mu \!\! &\equiv& \!\!
2 \bF_\mu(\bx)+ 5 \sigma_0^2 \dlx^2 \bF_\mu(x)\big|_{x=\bx}. 
%-----
\end{eqnarray}
Then we can write the one part in (\ref{Xmntr3}) as 
\begin{eqnarray}\label{msctakmehg}
&&
g^{\sigma\tau} g^{\rho\zeta}
\Big( 
\left< \bF_{\mu} \dl_\zeta \bga \dl_\sigma \bga \right> \left< \bF_{\nu} \dl_\rho  \bga \dl_\tau   \bga \right> 
- \, \left< \bF_{\mu} \right> g_{\zeta\sigma} \left< \bF_{\nu} \right> g_{\rho\tau}
\Big) \nn
%====================
[1.5mm]
&=& \, g^{\sigma\tau} g^{\rho\zeta} \, 
\Big\{ 
-g_{\zeta\sigma} g_{\rho\tau} \dl_\mu \phi(\theta) \dl_\nu \phi(\theta) \nn
%----------
[1.5mm] 
&&
\hspace{16mm}
+\,
{\cal F}_{0,\mu\zeta\sigma}
\Big( {\cal P}_{\nu\rho} \dl_\tau \phi(\theta) + {\cal P}_{\nu\tau}  \dl_\rho \phi(\theta) \Big) 
+
{\cal F}_{0,\nu\rho\tau}
\Big( {\cal P}_{\mu\zeta} \dl_\sigma \phi(\theta) + {\cal P}_{\mu\sigma}  \dl_\zeta \phi(\theta) \Big) \nn
%----------
[1.5mm]
&&
\hspace{16mm}
+\,
\Big( {\cal F}_{0,\mu\zeta\sigma} {\cal Q}_\nu + {\cal F}_{0,\nu\zeta\sigma} {\cal Q}_\mu \Big)
\dl_\rho \phi(\theta) \dl_\tau \phi(\theta) \nn
%----------
[1.5mm]
&& 
\hspace{16mm}
+\, 
\Big( {\cal P}_{\mu\zeta} \dl_\sigma \phi(\theta) + {\cal P}_{\mu\sigma}  \dl_\zeta \phi(\theta) \Big)
\Big( {\cal P}_{\nu\rho} \dl_\tau \phi(\theta)    + {\cal P}_{\nu\tau}    \dl_\rho \phi(\theta)  \Big) \nn
%----------
[1.5mm]
&&
\hspace{16mm}
+\,  {\cal Q}_\nu \dl_\rho \phi(\theta) \dl_\tau \phi(\theta) 
\Big( {\cal P}_{\mu\zeta} \dl_\sigma \phi(\theta) + {\cal P}_{\mu\sigma}  \dl_\zeta \phi(\theta) \Big) \nn
%----------
[1.5mm]
&&
\hspace{16mm}
%\left.
+\, {\cal Q}_\mu \dl_\zeta \phi(\theta) \dl_\sigma \phi(\theta) 
\Big( {\cal P}_{\nu\rho} \dl_\tau \phi(\theta)    + {\cal P}_{\nu\tau}  \dl_\rho \phi(\theta)    \Big) \nn
%----------
[1.5mm]
&&
\hspace{16mm}
+\, {\cal Q}_\mu {\cal Q}_\nu \dl_\rho \phi(\theta) \dl_\tau \phi(\theta) \dl_\zeta \phi(\theta) \dl_\sigma \phi(\theta) 
+ {\cal F}_{0,\mu\zeta\sigma} {\cal F}_{0,\nu\rho\tau}
\Big\}, 
\end{eqnarray} 
where we have used (\ref{Nmpnp}).  
Next, let us look at the rest part in (\ref{Xmntr3}). 
\newline

First, when $p$ is given as (\ref{dforgp}), we can write as
$\dl_{\sigma} g_{\mu\nu} = \dl_{\sigma} \dl_{\mu} \dl_{\nu} \phi$ according to (\ref{ggdrl2}). 
On the other hand, we can also write as 
$\dl_{\sigma} g_{\mu\nu} = - \left< \dl_{\sigma} \bga \dl_{\mu}\bga \dl_{\nu} \bga \right>$\footnote{
% ************** FOOT NOTE **************
We can calculate as
$
\dl_{\sigma} g_{\mu\nu} = 
- \left< \dl_{\sigma} \bga \dl_{\mu}\bga \dl_{\nu} \bga \right> 
+ \left< \left(\dl_{\sigma} \dl_{\mu} \bga \right) \dl_{\nu} \bga \right> 
+ \left<  \dl_{\mu} \bga \left(\dl_{\sigma} \dl_{\nu} \bga \right) \right> 
$. 
Using (\ref{FunZer}) and (\ref{ggdrl2}), we can see  
$
\left< \left(\dl_{\sigma} \dl_{\mu} \bga \right) \dl_{\nu} \bga \right>
=\left<  \dl_{\mu} \bga \left(\dl_{\sigma} \dl_{\nu} \bga \right) \right>=0
$.}.
Therefore, the following equality is held:\footnote{
In (\ref{resprmn}), 
using (\ref{metexpfm}) and the general relation $\dl_{\sigma}g^{\mu\nu}=-g^{\mu\alpha}g^{\nu\beta}\dl_{\sigma}g_{\alpha\beta}$,  
it is possible to write the front factor as 
$
g^{\sigma\tau}g^{\rho\zeta} \dl_{\rho} \dl_{\sigma} \dl_{\tau} \phi 
=  g^{\sigma\tau}g^{\rho\zeta} \dl_{\sigma} g_{\rho\tau}
= -\dl_\sigma g^{\sigma \zeta}
$.
% **************************************
} 
%==========
\begin{align}\label{frchgggpsi}
\left< \dl_\sigma \bga \dl_\mu \bga \dl_\nu \bga \right> = -\dl_\sigma \dl_\mu \dl_\nu \phi. 
\end{align}
Therefore, we can write the rest part as  
\begin{align}\label{resprmn}
g^{\sigma\tau}g^{\rho\zeta}
\left< \dl_{\rho}\bga\dl_{\sigma}\bga\dl_{\tau}\bga \right> \left< \bF_{\mu} \bF_{\nu}\dl_{\zeta}\bga \right>
= \,\,
-g^{\sigma\tau}g^{\rho\zeta} 
\left(\dl_{\rho} \dl_{\sigma} \dl_{\tau} \phi\right) \left< \bF_{\mu} \bF_{\nu}\dl_{\zeta}\bga \right>.
\end{align} 
Let us evaluate $\left< \bF_{\mu}\bF_{\nu} \, \dl_{\zeta}\bga \right>$ in the same manner with (\ref{FGGvan1}):
\begin{eqnarray}
\label{bFbFdgexpTR}
&& \!\!
\left< \bF_{\mu}\bF_{\nu} \, \dl_{\zeta}\bga \right> 
\nn
%==========
[1.5mm]
&&=
\hspace{0.0mm}
\Bigg< 
\left( \bF_\mu(\bx) + \dlx \bF_\mu(x)\big|_{x=\bx}(x-\bx) + \frac{1}{2}\dlx^2 \bF_\mu(x)\big|_{x=\bx}(x-\bx)^2 \right)
\nn 
%-----
[1.5mm]
&&
\hspace{4.5mm}
\times
\left(
\bF_\nu(\bx) + \dlx \bF_\nu(x)\big|_{x=\bx}(x-\bx) + \frac{1}{2}\dlx^2 \bF_\nu(x)\big|_{x=\bx}(x-\bx)^2 
\right)
\nn 
%-----
[1.5mm]
&&\hspace{4.5mm}
\times 
\left\{
\dl_\zeta \bga(x)\big|_{x=\bx} \left( 1 - \frac{(x-\bx)^2}{\sigma_0^2} \right) - \dlx \bF_\zeta(x)\big|_{x=\bx}(x-\bx)
\right\}
\Bigg>, 
\nn
%==========
[1.5mm]
&&=
-\sigma_0^2
\Big(
2 \dlx \bF_\mu(x)\big|_{x=\bx} \dlx \bF_\nu(x)\big|_{x=\bx} + 3 \sigma_0^2 \dlx^2 \bF_\mu(x)\big|_{x=\bx} \dlx^2 \bF_\nu(x)\big|_{x=\bx}
\Big)
\dl_\zeta \phi(\theta)
+{\cal F}_{1,\mu\nu\zeta}. \nn  
\end{eqnarray}
where we have defined the $\theta$-independent constant part in the one above as 
\begin{eqnarray}
\label{dfF1mnz}
{\cal F}_{1,\mu\nu\zeta}  
\!\!\! &\equiv& \!\!\!
-\dlx \bF_\zeta(x)\big|_{x=\bx} 
\Bigg\{
\sigma_0^2 \Big( \bF_\mu(\bx)  \dlx \bF_\nu(x)\big|_{x=\bx} + \bF_\nu(\bx) \dlx \bF_\mu(x)\big|_{x=\bx} \Big)
\nn 
%-----
&&
+\frac{3}{2}\sigma_0^4
\Big(
\dlx \bF_\mu(x)\big|_{x=\bx} \dlx^2 \bF_\nu(x)\big|_{x=\bx} + \dlx \bF_\nu(x)\big|_{x=\bx} \dlx^2 \bF_\mu(x)\big|_{x=\bx} 
\Big)
\Bigg\}. 
\end{eqnarray}
We can see ${\cal F}_{1,\mu\nu\zeta}={\cal F}_{1,\nu\mu\zeta}$.
Therefore,
\begin{eqnarray}\label{opricsif} 
\textrm{(\ref{resprmn})}  
\!\!\! &=& \!\!\!
\sigma_0^2 \, g^{\sigma\tau}g^{\rho\zeta} 
\left(\dl_{\rho} \dl_{\sigma} \dl_{\tau} \phi(\theta) \right)
\bigg\{
\Big(
2 \dlx \bF_\mu(x)\big|_{x=\bx} \dlx \bF_\nu(x)\big|_{x=\bx}
\nn
%-----
[1.5mm]
&&
\hspace{28.5mm} 
+
3 \sigma_0^2 \dlx^2 \bF_\mu(x)\big|_{x=\bx} \dlx^2 \bF_\nu(x)\big|_{x=\bx}
\Big)
\dl_\zeta \phi(\theta)+{\cal F}_{1,\mu\nu\zeta}  
\bigg\}. 
\end{eqnarray} 
\newline

We can write $R_{\mu\nu}$ in (\ref{Xmntr3}) just as $\textrm{(\ref{msctakmehg})}+\textrm{(\ref{opricsif})}$. 
Then, we can obtain the Ricci scalar, and can obtain the EH action described by $\theta^\mu$, $\bF_\mu$ and $\phi$. 
We write it formally as 
\begin{eqnarray}\label{RmnEHT}
S = \frac{1}{16\pi G_N'} \int d^D\theta \sqrt{-g} \, {\cal L}, \quad
{\cal L} = g^{\mu\nu}\, \big(\textrm{(\ref{msctakmehg})}+\textrm{(\ref{opricsif})}\big), 
\quad
\textrm{where $D=2$,}
\end{eqnarray} 
and $G_N'$ means the value of the Newton constant at the high energy region, therefore we consider the action above as 
the one at the ultra-violet fixed point. Actually, we perform the coarse-graining in the following sections, and the coarse-graining 
is the translation going from the high energy to low energy (generally speaking, coarse-graining is to look the low energy and long range behavior  
of the system by summarizing up the fine behaviors). Thus, we will determine $G_N'$ when we obtain the form of the above 
action at the low energy limit so that the part corresponding to the Newton constant at that low energy limit can be the $G_N$.

As such, although we have specified the Gaussian type of $p$ giving a two-dimensional space-time and expanded around $x=\bar{x}$, 
we have obtained the EH action in terms of the ingredients of the Fisher metric without the problem mentioned in the last of the introduction, 
which is one of results in this study.  Lastly we list the points in the rewriting of the EH action and the problems in that: 
\begin{itemize}
\item
We have performed the expansions around $\bx$ as in (\ref{bFmuTyl}) and (\ref{gGmuTyl}). 
At this time, for the form of $p$ we have taken specifically in (\ref{pAdS2Fmbt}), the expansions can stop at the second order as in (\ref{dlnxdlmbgvz}).  
\item
We have used the condition: $\int \! dx \, p(x,\theta) = 1$ as in (\ref{FunZer}).
\end{itemize}
As the problems in the rewritten EH action: 
\begin{itemize}
\item    
The metrices in this study are finally given in terms of the coordinates as (\ref{cogrgmn1}) and (\ref{costgmn2}).  
As a result, our action depends on the coordinates in complicated manner. Consequently, we cannot rewrite our action in the momentum space as we cannot perform 
the integral with regard to the coordinates in the Fourier transformation\footnote{
Heat-kernel method is often used to perform the coarse-graining in curved spaces 
(for one of reviews for this, see \cite{Vassilevich:2003xt}).}.  

\item
Since $g_{\mu\nu}$ and $g^{\mu\nu}$ are given by $\phi$, there is no quadratic term in the action. 
Therefore, there is no two-point correlated functions, and the perturbative analysis with the Wick contraction is unavailable.  
\end{itemize}

%================================================================================================================== 
\section{Coarse-grainings of the ingredients in our rewritten EH action} 
\label{Sec:Renomalization}
%================================================================================================================== 
 
In the sections so far, 
we have considered the Fisher metric given by $p=e^{-\bga}$, 
where $\bga(x,\theta) = \theta^\mu \bF_\mu (x) - \phi(\theta)$, 
and the components of those have been given in Sec.(\ref{Sec:EPO}). 
With those, we have obtained the rewritten EH action as in (\ref{RmnEHT}). 

In this section, 
we define our system in Sec.\ref{Sec:RenomaLaDAc}, 
then introducing the fixed-points,  
we give the transformation rules of the ingredients in our rewritten EH action under the coarse-graining and scale-down. 
In particular, the transformation rule of the Fisher metric is one of results in this study. 
(In what follows, if we say ``under the coarse-graining'', it means ``under the coarse-graining and scale-down''.)

%================================================================================================================== 
\subsection{Space we put our action} 
\label{Sec:RenomaLaDAc}
%================================================================================================================== 

We first define the space on which we put our action (\ref{RmnEHT}), which we refer to as $\Lambda$. 
We assume $\Lambda$ as a $D$-dimensional cubic lattice with the lattice spacing $1$ and the length of each side is $N$. 
Therefore, 
\begin{itemize}
\item the lattice points exist every lattice spacing $1$, 
\item the total number of the lattice points in $\Lambda$ is $N^D$,  
\end{itemize}
where $D$ is common in the one in Sec.\ref{Sec:What}.
We impose the periodic boundary condition in each direction, 
therefore 
the $\Lambda$ can be considered as a $D$-dimensional torus. 
Correspondingly, $N$ is assumed as even integers. 

We denote the lattice points $\theta$ in $\Lambda$ as $(\theta^1,\theta^2,\cdots \theta^D)$, 
where each component is integers satisfying $|\theta^i| \le N/2$. 
We consider $\phi(\theta)$ on each lattice point, where $\phi(\theta) \in \mathbb{R}$.

We have $\phi(\theta)$ exist on each lattice point of $\Lambda$, which is a curved space with the Fisher metric determined from $\phi(\theta)$. 
Therefore, in this study we are not considering $\phi(\theta)$ on some flat Euclidian space separately from $\Lambda$   
(in this case, using the information there, we will come to constitute the Fisher metric and EH action, 
but there is no grand for the connection between the theories in the flat Euclidian space and $\Lambda$),
but considering only $\Lambda$. 
($\bF_\mu(x)$ is considered a kind of parameter). 

In general, the systems in the statistical mechanics and field theories have degree of the freedom 
more than the Avogadro constant. Therefore $N$ is finally taken to infinity.

%================================================================================================================== 
\subsection{Coarse-graining of $\phi$, fundamental definition of renormalization transformation and important problem}
\label{SSec:RLS}
%================================================================================================================== 

In this subsection, we define the fundamental points of our renormalization transformation.  
\newline

\underline{\bf Definition of  ``scale-down'':}~
We first consider sufficiently large odd integers $L$, 
to which we refer as the ``renormalization scale''.  
$N$ is assumed to be multiples of $L$. 
Let us denote what we have mentioned now as 
\begin{itemize}
\item \textrm{$L$ is some odd integers to which we refer as the ``renormalization scale'',}
\item \textrm{$N = N' L$, where $N'$ is some constant even integers.}
\end{itemize}

Then, separately from $\Lambda$, 
we consider another $D$-dimensional cubic lattice with the length of each side is $N/L = N' \in$ even $\mathbb{Z}$, 
where its lattice spacing is $1$.  
We refer to this lattice space as $\Lambda'$. 
We consider that $\Lambda'$ is obtained from $\Lambda$ by dividing the each side by $L$, 
which we refer to as the ``scale-down''.
\newline
  
\underline{\bf Definition of  ``coarse-graining'':}~ 
For the lattice points $\eta=\left( \eta^1, \eta^1,\cdots, \eta^D \right)$ in $\Lambda'$ 
(We use $\eta^\mu$ as the notations of the coordinate in $\Lambda'$ as well as $\theta^\mu$ in $\Lambda$ in what follows.), 
we now define the region in $\Lambda$ that we refer to as $B_\eta$ as 
\begin{align}\label{DefBLT}
B_\eta = 
\left\{ 
\theta=\left( \theta^1, \theta^2,\cdots, \theta^D \right) \in \Lambda \, \bigg| \, \eta \in \Lambda' 
\,\,\,\, {\rm and} \,\,\, 
\left| \theta^i-L \eta^i \right| \le \frac{L-1}{2} \right\}. 
\end{align}   

Then if $\Lambda'$ can be obtained from $\Lambda$ through the scale-down by $1/L$, 
let us consider that the field $\phi'(\eta)$ in $\Lambda'$ is a mass of $\phi(\theta)$ in the region $B_\eta$ in $\Lambda$. 
This consideration leads to write the relation between $\phi'(\eta)$ and $\phi(\theta)$ as 
\begin{align}\label{CGO}
\phi'(\eta)
=\frac{1}{L^{\Delta}}\sum_{\theta \in B_\eta}\phi(\theta)
=\frac{1}{L^{\Delta}}\sum_{\theta \in B_0}\phi(L \eta  + \theta), 
\end{align}
where $0$ means zero vector in $\Lambda'$. We refer to this as  ``coarse-graining'' in what follows. 
\newline

%---
From (\ref{CGO}), we can get an interpretation that coarse-graining is the manipulation 
to summarize fine dynamics in some region to one local dynamics (thus 
coarse-graining is the transformation going from high-energy to low-energy). 
We fix $\Delta$'s concrete value in Sec.\ref{Sec:EHAofrt} 
such that the fixed-points which we introduce in the next subsection can exist.   
At this time,  anomalous dimension is  involved if we evaluate the contribution 
of the loop diagrams in the path-integral (\ref{RTHtoHd}), which shifts $\Delta$ to $\Delta-\eta$ ($\eta$ means anomalous dimensions). 
However, since the actual form of the action is given by very complicated one, 
we cannot evaluate  the contribution of the loop diagrams. In this sense, 
the quantum effect is not involved in the analysis in this study.

Combining the coarse-graining with the scale-down, we can interpret (\ref{CGO}) as the effective description 
when we look at the system further away. Therefore, we can interpret the coarse-graining as the manipulation 
to move from high-energy to low-energy. 
%--- 

The action $\int_{\Lambda'} d^D\eta \sqrt{-g'}\,{\cal L}'(\phi'(\eta))$ in the $\Lambda'$ system 
with a configuration of  $\phi'\equiv \{\phi'(\eta) \,\big|\, \eta \in \Lambda'\}$ can be defined 
from the ingredients in the $\Lambda$  system as
\begin{eqnarray}\label{RTHtoHd}
\exp \left[ -\int_{\Lambda'} d^D \eta \sqrt{-g'}\,{\cal L}'(\phi'(\eta)) \right] 
\!\! &=& \!\!
{\cal N}_0 \int {\cal D}\phi  
\left\{  
\prod_{\eta \in \Lambda'} \delta 
\left( \phi'(\eta) - \frac{1}{L^{\Delta}} \sum_{\theta \in B_\eta}\phi(\theta) \right) 
\right\} 
\nn
%-----
[1.5mm]
&& \!\!\!\! \times \exp \left[- \int_{\Lambda} d^D\theta \sqrt{-g}\,{\cal L}(\phi(\theta))\right],
\\
%-----
[1.5mm]
{\rm where} \quad
\int {\cal D} \phi \, (\cdots) \!\! &\equiv& \!\! 
\prod_{\theta \in \Lambda} \left( \int^\infty_{-\infty} d\phi(\theta) \right)(\cdots). 
\nonumber 
\end{eqnarray}
In the one above, 
$\phi(\theta)$ is real functions and ${\cal N}_0$ is a constant by which the r.h.s. can become unit when $\phi'=0$. 
Therefore, we can write ${\cal N}_0$ as
\begin{eqnarray}\label{defcalNu}
{\cal N}_0{}^{-1} 
=
\int {\cal D}\phi  
\prod_{\eta \in \Lambda'} \delta \left(\sum_{\theta \in B_\eta}\phi(\theta) \right)       
\exp \left[- \int_{\Lambda} d^D\theta \sqrt{-g}\, {\cal L}(\phi)\right].
\end{eqnarray}

(\ref{RTHtoHd}) is the definition of the ``renormalization transformation'',   
which is composed of the coarse-graining (\ref{CGO}) and scale-down
(If some coefficients, etc also get modification, exchanges concerning those are also included in (\ref{RTHtoHd})).

Since higher momenta are integrated out in the renormalization transformation, the renormalization transformation 
corresponds to enlarge the scale for our looking at the theory by renormalizing fine behaviors into ingredients in the theory.  

We cannot evaluate the concrete value of ${\cal N}_0$, since we do not know how to perform a path-integral for the action given as (\ref{RmnEHT}).
Since ${\cal N}_0$ will affect the cosmological constant, it follows that we cannot know how the cosmological constant is in our renomalization-flow. 

Lastly, we would like to stress the point we mentioned in Sec.\ref{Chap:Intro} and in the last of  Sec.\ref{Sec:What}.   
\newline

Let us bluffly write the $\phi$ in the path-integral (\ref{RTHtoHd}) and (\ref{defcalNu}) like 
$\phi=\textrm{(\ref{pAdS2FmPbt})}+\cdots$, where $``\cdots"$ means the configuration 
arbitrarily generated from path-integral. 
%---
From this, we can see that $\phi$ will be no longer in the form  (\ref{pAdS2FmPbt}) 
in the pass-integral for the the part $``\cdots"$. Correspondingly, we can see that $p$ 
in (\ref{dforgp}) no longer satisfy (\ref{sump1}) as it gets deviated from the Gaussian form 
(\ref{pAdS2Fmbt}).

This point is very critical in the study of the Fisher metric. 
However it is a fact that we should involve the quantum effect in our study. 
Therefore, accepting that (\ref{sump1}) is kept in the classical situation but is broken 
at the quantum level, we consider the quantum effect (the part $``\cdots"$ (deviated part)) of $\phi$.   
At this time, we consider that some constraints and causal relations held at the classical level usually get broken at the quantum level for the quantum uncertainty.  

The problem we mentioned above has arisen from our considering of the fundamental quantity $p$ 
in the Fisher metric using a field $\phi$, however this problem may be the problem always existing 
as long as $p$ is defined using some fields (and probably this problem may exist even in the quantum mechanics). 
If so, this problem would be very serious problem in considering the Fisher metric in physics. 

In order to resolve this problem we may ought to  consider further  constraints 
by further inserting delta functions in the path-integral, however anyway 
becoming able to carry out the analysis for the quantum effect of Fisher metric 
seems very hard. (If it comes to this study, looking at (\ref{pAdS2FmPbt}) and (\ref{pAdS2Fmbt}) 
we can see that there is no d.o.f. anymore to involve quantum fluctuation with 
(\ref{sump1}) kept. This is because we can see there is no d.o.f. even adding 
constants in (\ref{pAdS2FmPbt}) without (\ref{sump1}) broken.) Resolving this 
problem would be an important problem if we consider the Fisher metric in physics.
\newline

Indeed, there is one more point we should take care. It is that  
there is no longer guarantee that the (\ref{sump1}) is being satisfied 
by the $p$ given by the $\phi'(\eta)$ even at the classical level, 
where $\phi'(\eta)$ means the ones obtained after the coarse-gainings 
according to (\ref{CGO}). 
This is because the normalization parameter 
$L^{-\Delta}$ in (\ref{CGO}) is to be determined 
irrelevantly with (\ref{sump1}). (Namely the $p$ with the $\phi(\theta)$  can satisfy (\ref{sump1}) by definition, 
however there is no guarantee that $p$ with the $\phi(\eta)$ can satisfy (\ref{sump1}).)   

However, we may be able to consider this problem no critical in this study. 
This is because (\ref{CGO}) is just to prescribe how so-called mean-$\phi$'s 
after the coarse-grainings (\ref{CGO}) are looked, and the $\phi$'s at the stage 
with no coarse-grainings performed are always satisfying (\ref{sump1}) by definition.  
Therefore, this problem can be considered just as an apparent problem.  
\newline

Summarizing the ones above, 
\begin{itemize}
\item   
at the classical level, 
we consider $\phi'$'s prescribed as (\ref{CGO}), which no longer satisfy (\ref{sump1}). 
However it would not become problems 
by considering that it is just an apparent problem as (\ref{sump1}) are being satisfied at the stage with no coarse-grainings. 
\item   
at the quantum level, we consider the fluctuation (the quantum effect) on $\phi$, which breaks (\ref{sump1}) from the first stage where no coarse-grainings (high-energy region).
\end{itemize}

%================================================================================================================== 
\subsection{Fixed-points}
\label{SSec:FPEP}
%================================================================================================================== 

Denoting $\int_\Lambda d^D\theta \sqrt{-g}\, {\cal L}(\phi)$ and $\int_{\Lambda'} d^D\eta \sqrt{-g'}\,{\cal L}'(\phi')$ in (\ref{RTHtoHd}) 
as $S^{(i)}$ and $S^{(i+1)}$, let us consider a sequence, $S^{(0)} \to S^{(1)} \to \cdots \to S^{(i)} \to S^{(i+1)} \to \cdots$, 
generated by $S^{(i+1)}={\cal R} \cdot S^{(i)}$, 
where ${\cal R}$ means the renormalization transformation prescribed in (\ref{RTHtoHd}).

Now we consider the action invariant with respect to the renormalization transformation ${\cal R}$.  
Then, denoting the invariant action as $S^*$, we can write as
\begin{align}
{\cal R} \cdot S^* = S^*.  
\end{align}
We refer to $S^*$ as the ``fixed-point''. 
One can reach the fixed-point up to 
the values of $\Delta$ in (\ref{CGO}). 
In the following  sections in this study, 
we obtain the $\Delta$ corresponding to the fixed-points.

%================================================================================================================== 
\subsection{Coarse-gainings other than $\phi$}
\label{SSec:Mean}
%================================================================================================================== 

In this subsection, we give the transformation rules under the coarse-grainings for the ingredients, 
$\theta^\mu$, $\dl^\mu$, $\bF_\mu$ and $g_{\mu\nu}$ and so on. 
In particular, determining the transformation rule of the Fisher metric under the coarse-graining is one of the results in this study.
\newline

Regarding $\theta^\mu$, those play the role of the coordinates in $\Lambda$, 
which are not the target of the coarse-graining. 
As mentioned in Sec.\ref{SSec:RLS}, those just get the scale-down in one renormalization transformation as 
\begin{eqnarray}\label{NmpnpTh}
\eta^\mu \!\!&=&\!\! {\theta^\mu}/{L}, \quad 
\eta_\mu = {\theta_\mu}/{L}, 
\end{eqnarray} 
where $L$, $\eta^\mu$ and $\theta^\mu$ are defined in Sec.\ref{SSec:RLS}, 
and $\eta^\mu$ and $\theta^\mu$ represent the coordinates after and before the scale-down, 
which we can denote as  
\begin{eqnarray}\label{ntnwuts}
\eta^\mu \in \Lambda', \quad 
\theta^\mu \in \Lambda.
\end{eqnarray}
We use those notations through this paper.  
Note that since the scale-down is the transformation irrelevant with the indices $\mu$ and $\nu$, 
there is no difference originating in contravariant and covariant vectors. 
Along with (\ref{NmpnpTh}), in one scale-down, $\dl/\dl \eta^\mu$ and $\dl/\dl \eta_\mu$ are transformed as
\begin{eqnarray}\label{Nmpnpdl}
\frac{\dl}{\dl \eta^\mu} \!\!&=&\!\! L \frac{\dl}{\dl \theta^\mu}, \quad 
\frac{\dl}{\dl \eta_\mu} = L \frac{\dl}{\dl \theta_\mu}.
\end{eqnarray} 

Next, since each term of (\ref{dforggm}) are transformed in the same manner under the coarse-grainings, 
if $\phi(\theta)$ and ${\dl}/{\dl \theta^\mu}$ are transformed as (\ref{CGO}) and (\ref{Nmpnpdl}), $\bF_\mu(x)$ in one coarse-graining is transformed as
\begin{align}\label{NmpnpF} 
\bF_\mu'(x) = L^{-\Delta+1}\sum_{\theta \in B_\eta}\bF_\mu(x)=L^{D-\Delta+1} \bF_\mu(x).     
\end{align} 
In the one above, 
$\mu$ in the l.h.s. and r.h.s. indicate the coordinates $\eta^\mu$ and $\theta^\mu$ respectively, 
and $x$ is that appearing in Sec.\ref{Sec:What}. 
We have evaluated the summation 
by using the fact that $\bF_\mu(x)$ is independent of $\theta^\mu$ and the number of the lattice points in $B_\eta$ is $L^D$,

$\xi^\mu \, (= L^{-1} \, \eta^\mu)$, 
$\dl/\dl \xi^\mu \, (= L \, \dl/\dl  \eta^\mu)$ 
and 
$\bF_\mu''(x) \,(= L^{\left(D-\Delta+1\right)} \bF'_\mu(x))$, 
ones obtained by performing the scale-down twice, can be written as
\begin{eqnarray}\label{xidxibfddtw} 
\xi^\mu = \theta^\mu/L^2, \quad
\frac{\dl}{\dl \xi^\mu} = L^2 \frac{\dl}{\dl \theta^\mu},  \quad
\bF_\mu''(x) = L^{2\left(D-\Delta+1\right)} \bF_\mu(x),
\end{eqnarray}
where $\xi^\mu$ mean the coordinates with the scale-down twice from $\Lambda''$ (so, we can write as  $\xi^\mu \in \Lambda''$).

When $\bF_\mu(x)$ are transformed as (\ref{NmpnpF}), 
${\cal F}_{0,\mu\zeta\sigma}$, ${\cal P}_\mu$, ${\cal Q}_\mu$ and ${\cal F}_{1,\mu\nu\zeta}$ given 
in (\ref{thinptc}), (\ref{dfPmr}), (\ref{dfQmr}) and (\ref{dfF1mnz}) 
are transformed by the following manners:
\begin{eqnarray}
\label{tfrFcaqo} 
{\cal F}'_{0,\mu\zeta\sigma} \!\! &=& \!\! L^{3\left(D-\Delta+1\right)} \, {\cal F}_{0,\mu\zeta\sigma}, \qquad
{\cal P}'_\mu \,=\, L^{2\left(D-\Delta+1\right)} \, {\cal P}_\mu, \\ 
[1.5mm] 
\label{tfrQcaqo} 
{\cal Q}'_\mu \!\! &=& \!\! L^{D-\Delta+1} \, {\cal Q}_\mu, \qquad \hspace{3.5mm}
{\cal F}'_{1,\mu\nu\zeta} \,=\, L^{3\left(D-\Delta+1\right)} \, {\cal F}_{1,\mu\nu\zeta}.
\end{eqnarray} 
\newline 

When $\phi$ gets the coarse-graining  as in (\ref{CGO}), 
the Fisher metric gets the coarse-graining as can be seen from its definitions given in Sec.\ref{Sec:What} and \ref{Sec:EPG}. 
We have given the definitions of the Fisher metric in several ways.   
The coarse-grained Fisher metrics obtained from those should agree each other. 
For example, the results obtained from (\ref{DefFishmet1}) and (\ref{metexpfm}) should agree.  
However, as shown in what follows, those do not agree in fact.

If we follow (\ref{metexpfm}), we can write the Fisher metric after one coarse-graining as
\begin{eqnarray}\label{Nmpnpg} 
g'_{\mu\nu}(\eta) 
= \left< \frac{\dl}{\dl \eta^\mu} \frac{\dl}{\dl \eta^\nu} \phi'(\eta) \right> 
\!\! &=& \!\!  \frac{\dl}{\dl \eta^\mu} \frac{\dl}{\dl \eta^\nu} \phi'(\eta) \left<1\right> \nn
[1.5mm]
\!\! &=& \!\! 
L^{-\Delta+2} \frac{\dl}{\dl \theta^\mu} \frac{\dl}{\dl \theta^\nu} 
\sum_{\theta \in B_0}\phi(L \eta  + \theta) \, \int dx \, e^{-\bga'(x, \eta)}, 
\end{eqnarray} 
where $\langle \cdots \rangle$ is the statistical average with regard to $x$ as defined in (\ref{ladddrlstav}).

Let us evaluate $\int dx \, e^{-\bga'(x, \eta)}$. We can write $\bga'(x,\eta)$ as
\begin{eqnarray}\label{Nmpnpbga} 
\bga'(x,\eta) 
\!\!&=&\!\! 
- \eta^\mu \, \bF_\mu'(x) + \phi' (\eta) \nn 
%---
[1.5mm]
\!\!&=&\!\! 
\frac{1}{L^{\Delta}} \sum_{\theta \in B_\eta} \left( - \eta^\mu L \, \bF_\mu(x)
+ \phi(L\eta+\theta)\right) \nn
%---
[1.5mm]
\!\!&=&\!\! 
\frac{1}{L^{\Delta}} \sum_{\theta \in B_\eta} \left( - \eta^\mu L \, \bF_\mu(x)
+ \frac{1}{2} \ln \left[ -\frac{\pi}{L\eta^2+\theta^2} \right] - \frac{(L\eta^1+\theta^1)^2}{4(L\eta^2+\theta^2)} \right), 
\end{eqnarray} 
where we have used (\ref{pAdS2Fmbt}).
In order to make our analysis possible, we approximately assume that $\phi(\theta)$ is unique in each region $B_\eta$. 
More concretely, this is an approximation regarding 
$\phi(L\eta+\theta) = \phi(L\eta) + \phi'(L\eta)\theta + \frac{1}{2}\phi''(L\eta)(\theta)^2+\cdots$ as $ \phi(L\eta)$ in each region $B_\eta$. 
Then, we can write the Fisher metric (\ref{Nmpnpg}) as
\begin{eqnarray}
\label{Cgpnpbga1} 
g'_{\mu\nu}(\eta) \!\! &\sim& \!\! 
L^{D-\Delta+2} \, \frac{\dl}{\dl \theta^\mu} \frac{\dl}{\dl \theta^\nu} \phi(\theta)  
\int dx \, e^{-\bga'(x, \eta)}, \\
%%----------
[1.5mm]
\label{Cgpnpbga2a}
{\rm where}\quad
\bga'(x,\eta) 
\!\! &\sim& \!\! 
L^{D-\Delta} \left( 
- L\eta^\mu \, \bF_\mu(x) 
+ \frac{1}{2} \ln \left[ -\frac{\pi}{L\eta^2} \right] - \frac{(L\eta^1)^2}{4L\eta^2}  
\right) \nn
%----------   
[1.5mm]
\label{Cgpnpbga2c}
\!\! &=& \!\!  
\frac{L^{D-\Delta}}{2\sigma_0^2}\left( \left(x-\bx \right)^2+\sigma_0^2 \ln \left[2\pi \sigma_0^2\right] \right),
\end{eqnarray} 
``$\sim$'' means we have used the approximation mentioned between (\ref{Nmpnpbga}) and (\ref{Cgpnpbga1}) 
(it can be written as ${\cal O}(L)$ as the amount having been truncated, which can be gotten by expanding with regard to  $\theta^{1,2}$ in (\ref{Nmpnpbga}) to the first-order), 
and using $L(\eta^1,\eta^2)=(\theta^1,\theta^2)$ we have exchanged $\eta^\mu$ to $\theta^\mu$, 
then rewritten with (\ref{pAdS2FmFbt}) and (\ref{pAdS2FmTbt}).

Then, $g_{11}$ and $g_{12}$ calculated based on (\ref{Cgpnpbga1}) can agree 
with those calculated based on (\ref{DefFishmet1}) using the same $\bga'$ in (\ref{Cgpnpbga2a}), 
however $g_{22}$ cannot agree. 
Its reason is that $\int dx \, e^{-\bga'(x, \eta)}$ depends on the coordinates as
\begin{eqnarray}
\label{brconty}
\int_{-\infty}^\infty dx \, e^{-\bga'(x, \eta)} \!\! &=& \!\! 
\sqrt{
\frac{1}{L^{D-\Delta}}
\left(-\frac{\theta^2}{\pi}\right)^{-1+L^{D-\Delta}}
},
\end{eqnarray} 
where we have used 
\begin{eqnarray}
\bga'(x,\eta) = 
-\frac{L^{D-\Delta}}{4\theta^2}
\left((\theta^1+2x \theta^2)^2-2\theta^2\ln\left[-\frac{\pi}{\theta^2}\right]\right). 
\end{eqnarray}
This can be equivalently obtained from (\ref{Cgpnpbga2c}).
Therefore, 
$\dl_\mu \left(\int dx \, e^{-\bga'(x, \eta)}\right) =0$ is not held. 
This is the condition written under (\ref{octprorw}), and due to this, 
we cannot rewrite (\ref{DefFishmet1}) to (\ref{DefFishmet3}). 
If $L^{D-\Delta}=1$, (\ref{brconty}) can be unit, and at this time we can confirm $g_{22}$ can also agree. 
\newline

As such, how to determine the coarse-grained Fisher metric is a problem.   
Although we can make logic for this variously. 
we here would like to give the one we can organize consistently to the end, 
which starts with the notion that 
if the coarse-grained scalar field is given as (\ref{CGO}), 
also for the coarse-graining of the tensor field, we may write as 
\begin{eqnarray}\label{cggmuiln1}
g'{}^{\mu\nu}(\eta) 
= \sum_{\theta \in B_\eta}g^{\mu\nu}(L\eta+\theta) \sim L^{D - \Delta_g} g^{\mu\nu}(L\eta) 
= L^{D - \Delta_g}  g^{\mu\nu}(\theta),
\end{eqnarray} 
where $\Delta_g$ is intended to play the same role with $\Delta$ in (\ref{CGO}), 
and ``$\sim$'' has the same intention with the one in (\ref{Cgpnpbga1}).

Since the coarse-graining is irrelevant with the contravariant and covariant vectors,
if we can write like (\ref{cggmuiln1}), we can also write as
\begin{eqnarray}\label{cggmuiln2}
g'_{\mu\nu}(\eta) \sim L^{D - \Delta_g} \, g_{\mu\nu}(L\eta) 
=  L^{D - \Delta_g} \,  g_{\mu\nu}(\theta).
\end{eqnarray}

Then, for some vectors $V^\mu(\eta)$ in $\Lambda'$, 
we consider a relation: 
\begin{eqnarray}
V^\mu(\eta) = g'{}^{\mu\nu}(\eta) \, g'_{\nu\lambda}(\eta) \, V^\lambda(\eta). 
\end{eqnarray} 
The one above is a relation based on the fact that metrices in $\Lambda'$ are given by $g'{}^{\mu\nu}(\eta)$ and $g'_{\nu\lambda}(\eta)$.
Then, since $g^{\mu\nu}(L\eta) \, g_{\nu\lambda}(L\eta)=\delta^\mu_\lambda$, 
\begin{eqnarray}\label{vtrtr}
V^\mu(\eta)  = L^{2(D -\Delta_g)} \, V^\mu(\eta). 
\end{eqnarray}
Therefore, $\Delta_g = D$ is led, 
and the transformation rule for $n$ times coarse-grainings is determined from (\ref{cggmuiln1}) and (\ref{cggmuiln2}) as
\begin{eqnarray}
\label{cogrgmn1}   
&&
\bullet \quad 
g^{(n)}{}^{\mu\nu}(\zeta) 
\, = \,
2L^{2n} \, 
\left(
\begin{array}{cc}
\displaystyle (\zeta^1)^2-\zeta^2/L^n & \displaystyle \zeta^1 \zeta^2 \\
\displaystyle \zeta^1 \zeta^2         & \displaystyle (\zeta^2)^2
\end{array}
\right)
\equiv
L^{2n}\,\mathfrak{g}^{(n)}{}^{\mu\nu}(\zeta),
\\ 
%==========  
[1.5mm]
\label{costgmn2} 
&&
\bullet \quad
\hspace{4mm} 
g^{(n)}_{\mu\nu}(\zeta) 
\, = \,
\frac{1}{2L^n}
\left( 
\begin{array}{cc} 
\displaystyle -\frac{1}{\zeta^2}                        & \displaystyle \frac{\zeta^1}{(\zeta^2)^2} \\
\displaystyle \displaystyle \frac{\zeta^1}{(\zeta^2)^2} & \displaystyle \frac{-(\zeta^1)^2+\zeta^2/L^n}{(\zeta^2)^3}
\end{array}
\right)
\equiv L^{-n} \,\mathfrak{g}^{(n)}_{\mu\nu}(\zeta), 
\end{eqnarray}
where the superscripts ``$(n)$'' ($n=0,1,2,\cdots$) mean  
the number of the coarse-graining those got, 
and $\zeta^\mu$ mean the coordinates with $n$ times scale-down.

%==================================================================================================================  
\section{Coarse-graining and fixed-point of our rewritten EH action}
\label{Sec:EHAofrt}
%================================================================================================================== 
  
In Sec.\ref{Sec:EPG} and \ref{Sec:FM}, 
we have considered the Fisher metric, $p=e^{-\left( \theta^\mu \bF_\mu (x) - \phi(\theta) \right)}$ 
(as for the components of those, see Sec.(\ref{Sec:EPO})), and have obtained the rewritten EH action as in (\ref{RmnEHT}). 
In Sec.\ref{Sec:Renomalization}, introducing the fixed-points,  
we have given the transformation rules of the ingredients in our rewritten EH action under the coarse-graining. 
In this section, we perform the coarse-graining toward our rewritten EH action (\ref{RmnEHT}), 
then examine the fixed-points. 
\newline

Using the transformation rules (\ref{NmpnpF}), (\ref{xidxibfddtw}), (\ref{tfrFcaqo}), (\ref{tfrQcaqo}), (\ref{cogrgmn1}) and (\ref{costgmn2}), 
we perform the coarse-graining toward our rewritten EH action (\ref{RmnEHT}). 
We here give the transformation rule of $\phi(\theta)$ we employ based on (\ref{CGO}) as 
\begin{eqnarray}\label{cgoac}
\phi'(\eta) \sim L^{D-\Delta} \phi(\theta), 
\end{eqnarray} 
where ``$\sim$'' has the same intention with the one in (\ref{Cgpnpbga1}). 
Summarizing the manipulation we perform the coarse-grainings $n$ times from $S^{(0)}$ on $\Lambda^{(0)}$ to $S^{(n)}$ on $\Lambda^{(n)}$, 
\begin{eqnarray}\label{bklfd}
\theta^\mu &\to& L^n \zeta^\mu, \quad 
\theta_\mu \to L^n \zeta_\mu
\nn 
%----- 
[1.5mm]
\frac{\dl}{\dl \theta^\mu} \!\!&\to&\!\! \frac{1}{L^n} \frac{\dl}{\dl \zeta^\mu}, \quad 
\frac{\dl}{\dl \theta_\mu} \to \frac{1}{L^n} \frac{\dl}{\dl \zeta_\mu},\nn
%-----
[1.5mm] 
\label{phzlphn} 
\phi^{(0)}(\theta) &\to& L^{-n(D-\Delta)} \phi^{(n)}(\zeta),\\
%-----
[1.5mm]
\bF^{(0)}_\mu(x) \!\!&\to&\!\! L^{-n(D-\Delta+1)}\bF_\mu^{(n)}(x),\nn     
%-----
[1.5mm]
{\cal F}^{(0)}_{0,\mu\zeta\sigma} \!\! &\to& \!\! L^{-3n\left(D-\Delta+1\right)} \, {\cal F}^{(n)}_{0,\mu\zeta\sigma}, \qquad
{\cal P}^{(0)}_\mu \,\to\, L^{-2n\left(D-\Delta+1\right)} \,{\cal P}^{(n)}_\mu , \nn
%-----
[1.5mm] 
{\cal Q}^{(0)}_\mu \!\! &\to& \!\! L^{-n(D-\Delta+1)} \,{\cal Q}^{(n)}_\mu , \qquad \hspace{3.5mm}
{\cal F}^{(0)}_{1,\mu\nu\zeta} \,\to\, L^{-3n\left(D-\Delta+1\right)} \,{\cal F}^{(n)}_{1,\mu\nu\zeta},\nn
%-----
[1.5mm] 
\label{gcgrsn}
g^{(0)}{}^{\mu\nu}(\theta) \!\! &\to& \!\! g^{(n)}{}^{\mu\nu}(\zeta), \quad 
g^{(0)}_{\mu\nu}(\theta) \,\to\, g^{(n)}_{\mu\nu}(\zeta),\\
%----- 
[1.5mm] 
\label{sgcgrsn}
\sigma_0^2  \!\! &\to& \!\! \sigma_0^2 /L^{2n},
\end{eqnarray} 
where the reason for (\ref{sgcgrsn}) is given at (\ref{sgcgne}) 
since we would like to determine it after $\Delta$ is determined. 
\newline

Since $L$-dependences remain in $g^{(1)}{}^{\mu\nu}(\eta)$ and $ g^{(1)}_{\mu\nu}(\eta)$   
as can be seen in (\ref{cogrgmn1}) and (\ref{costgmn2}), 
%we may consider we should consider 
it may be considered that we should consider not (\ref{gcgrsn}) but (\ref{gcgrsnac}).  
%---
However, since the necessary condition as the metrices, 
%that's the each one is 
which is that each one is
inverse matrix for each other, 
is held in $g^{(1)}{}^{\mu\nu}(\eta)$ and $ g^{(1)}_{\mu\nu}(\eta)$, 
we consider the replacement in terms of $g^{(1)}{}^{\mu\nu}(\eta)$ and $ g^{(1)}_{\mu\nu}(\eta)$ as in (\ref{gcgrsn}). 
%---
In Appendix.\ref{App:cgrmtg} , we note how the situation will be if we consider (\ref{gcgrsnac}). 
%--- 

Depending on either of those, $\Delta$ will be different as in (\ref{delfx}) and (\ref{delfxac}), 
and in the case of (\ref{delfx}), 
the relations of (\ref{DefFishmet1}) and (\ref{metexpfm}) can be held 
for the coarse-grained $\phi^{(n)}(\zeta)$ and $g^{(n)}{}^{\mu\nu}(\zeta)$, $ g^{(n)}_{\mu\nu}(\zeta)$ for arbitrary $n$. 
For concrete things for this, see the last of this section.  
%--- 

This result means that the formulation system of the Fisher metric 
(this terminology means that the Fisher metric is given by $\phi^{(n)}(\zeta)$)  
can be held at the arbitrary $n$ times coarse-grainings, 
however this can hold or not in the process of the coarse-graining is highly nontrivial.  
This is because the coarse-graining is performed independently of the formulation system of the Fisher metric, 
and $\Delta$ is determined irrelevantly with the formulation system of the Fisher metric.

As shown in (\ref{delfx}), $\Delta$ will be determined to the proper value if we consider (\ref{gcgrsn}). 
However, currently we have no idea of what $\Delta$ can be determined to the proper is accident or not.
\newline

We can obtain our rewritten EH action with $n$ times coarse-grainings as 
\begin{eqnarray}\label{rehacg}
S^{(n)}(\zeta)   
= 
\frac{L^{nD}}{16\pi G_N'} \int d^D\zeta \sqrt{-g^{(n)}\left( \zeta \right)} \, {\cal L}^{(n)}\left(\zeta \right).  
\end{eqnarray}  
where $S^{(n)}$ is the theory on $\Lambda^{(n)}$ 
with the coordinate $\zeta^\mu = \theta^\mu / L^n$ for any $n$ as defined under (\ref{cogrgmn1}).
$L^{nD}$ comes from $ d ^D \theta \sqrt{-g(\theta)}$, 
and  ${\cal L}^{(n)} \left(\zeta \right)$ is given as  
%******************************************
\allowdisplaybreaks
%****************************************** 
\begin{eqnarray}\label{rehacg37}
{\cal L}^{(n)}(\zeta) 
&=& 
\hspace{4mm}
g^{(n)}{}^{\sigma\tau}\left(\zeta \right)
g^{(n)}{}^{\rho\zeta}\left(\zeta \right)
g^{(n)}{}^{\mu\nu}\left(\zeta \right) \,  
\bigg\{ \nn
%----------
[1.5mm] 
&&
- \,
L^{-2n(D-\Delta+1)}
g^{(n)}_{\zeta\sigma}\left(\zeta \right) \, 
g^{(n)}_{\rho\tau}\left(\zeta \right) \, 
\dl_\mu \phi^{(n)}\left(\zeta \right)\, \dl_\nu \phi^{(n)}\left(\zeta \right) \nn
%----------
[1.5mm] 
&&
+\,
L^{-6n(D-\Delta+1)} \,
{\cal F}^{(n)}_{0,\mu\zeta\sigma}
\Big( {\cal P}^{(n)}_{\nu\rho} \dl_\tau \phi^{(n)}\left(\zeta \right)+ {\cal P}^{(n)}_{\nu\tau} \dl_\rho \phi^{(n)}\left(\zeta \right)\Big) \nn 
%----------
[1.5mm] 
&&
+\,
L^{-6n(D-\Delta+1)} \,
{\cal F}^{(n)}_{0,\nu\rho\tau}
\Big( {\cal P}^{(n)}_{\mu\zeta} \dl_\sigma \phi^{(n)}\left(\zeta \right)+ {\cal P}^{(n)}_{\mu\sigma} \dl_\zeta \phi^{(n)}\left(\zeta \right)\Big) \nn
%----------
[1.5mm]
&&
+\,
L^{-6n(D-\Delta+1)} \,
\Big( {\cal F}^{(n)}_{0,\mu\zeta\sigma} {\cal Q}^{(n)}_\nu + {\cal F}^{(n)}_{0,\nu\zeta\sigma} {\cal Q}^{(n)}_\mu \Big)
\dl_\rho \phi^{(n)}\left(\zeta \right)\dl_\tau \phi^{(n)}\left(\zeta \right)\nn
%----------
[1.5mm]
&& 
+\, 
L^{-6n(D-\Delta+1)} \,
\Big( {\cal P}^{(n)}_{\mu\zeta} \dl_\sigma \phi^{(n)}\left(\zeta \right)+ {\cal P}^{(n)}_{\mu\sigma}  \dl_\zeta \phi^{(n)}\left(\zeta \right)\Big)
\Big( {\cal P}^{(n)}_{\nu\rho} \dl_\tau \phi^{(n)}\left(\zeta \right)   + {\cal P}^{(n)}_{\nu\tau}    \dl_\rho \phi^{(n)}\left(\zeta \right) \Big) \nn
%----------
[1.5mm]
&&
+\, 
L^{-6n(D-\Delta+1)} \,
 {\cal Q}^{(n)}_\nu \dl_\rho \phi^{(n)}\left(\zeta \right)\dl_\tau \phi^{(n)}\left(\zeta \right)
\Big( {\cal P}^{(n)}_{\mu\zeta} \dl_\sigma \phi^{(n)}\left(\zeta \right)+ {\cal P}^{(n)}_{\mu\sigma}  \dl_\zeta \phi^{(n)}\left(\zeta \right)\Big) \nn
%----------
[1.5mm]
&&
+\, 
L^{-6n(D-\Delta+1)} \,
{\cal Q}^{(n)}_\mu \dl_\zeta \phi^{(n)}\left(\zeta \right)\dl_\sigma \phi^{(n)}\left(\zeta \right)
\Big( {\cal P}^{(n)}_{\nu\rho} \dl_\tau \phi^{(n)}\left(\zeta \right)   + {\cal P}^{(n)}_{\nu\tau}  \dl_\rho \phi^{(n)}\left(\zeta \right) \Big) \nn
%----------
[1.5mm]
&& 
+\, 
L^{-6n(D-\Delta+1)} \,
{\cal Q}^{(n)}_\mu {\cal Q}^{(n)}_\nu 
\dl_\rho \phi^{(n)}\left(\zeta \right) \dl_\tau \phi^{(n)} \left(\zeta \right) 
\dl_\zeta \phi^{(n)}\left(\zeta \right)\dl_\sigma \phi^{(n)}\left(\zeta \right)\nn
%----------
[1.5mm]
&&
+\,  
L^{-6n(D-\Delta+1)} \,
{\cal F}^{(n)}_{0,\mu\zeta\sigma} {\cal F}^{(n)}_{0,\nu\rho\tau}
\bigg\}
\nn
%----------
[1.5mm]
&&
+\,\sigma_0^2 \, 
g^{(n)}{}^{\sigma\tau} g^{(n)}{}^{\rho\zeta}  
\left(\dl_{\rho} \dl_{\sigma} \dl_{\tau} \phi^{(n)}\left(\zeta \right)\right)
\bigg\{
\nn
%-----
[1.5mm]
&&
\hspace{4mm}
\Big(
\hspace{4mm}
2 L^{-4n(D-\Delta+2)} \, \dlx \bF^{(n)}_\mu(x)\big|_{x=\bx} \dlx \bF^{(n)}_\nu(x)\big|_{x=\bx}
\nn
%-----
[1.5mm]
&&
\hspace{7mm}
+\,3 L^{-2n(2D-2\Delta+5)} \, \sigma_0^2 \dlx^2 \bF^{(n)}_\mu(x)\big|_{x=\bx} \dlx^2 \bF^{(n)}_\nu(x)\big|_{x=\bx}
\Big)
\dl_\zeta \phi^{(n)}\left(\zeta \right)
\nn
%-----
[1.5mm] 
&&
\hspace{4mm}
+ \,L^{-4n(D-\Delta+2)} \, {\cal F}^{(n)}_{1,\mu\nu\zeta}  
\bigg\}.
\end{eqnarray} 
In the one above, $\dl_\mu=\dl/\dl \zeta^\mu$.

In $S^{(n)}$ above, we can see there appear four kinds of the exponents, 
which we express as $\kappa_{1,2,3,4}$ as
\begin{eqnarray}
\label{kdvpvdc}
            \kappa_1 \!\! &=& \!\! {D}-2(D-\Delta+1), \\ 
[1.5mm] \kappa_2 \!\! &=& \!\! {D}-6(D-\Delta+1), \\
[1.5mm] \kappa_3 \!\! &=& \!\! {D}-2(2D-2\Delta+5), \\
[1.5mm] \kappa_4 \!\! &=& \!\! {D}-4(D-\Delta+2),
\end{eqnarray}
where in the value of $\kappa_1$, 
we have taken into account of the two facts:  
1) $L^n \, \mathfrak{g}^{(n)}_{\delta\epsilon}(\zeta) \, \mathfrak{g}^{(n)}{}^{\epsilon\zeta}(\zeta)=\delta^\delta_\zeta$,  
2) We later take the contraction as mentioned under (\ref{njomzkmtr}). 
Then, if we take $\Delta$ such that 
\begin{enumerate}
\renewcommand{\labelenumi}{\alph{enumi}).}
\item 
$\kappa_1$ vanishes; $\Delta=\frac{2+D}{2}$, which leads to 
$\kappa_2=-2D$, 
$\kappa_3=-(D+6)$, 
$\kappa_4=-(D+4)$,
\item 
$\kappa_2$ vanishes; $\Delta=\frac{6+5D}{6}$, which leads to 
$\kappa_1=\frac{2D}{3}$, 
$\kappa_3=\frac{D-18}{3}$, 
$\kappa_4=\frac{D-12}{3}$,
\item 
$\kappa_3$ vanishes; 
$\Delta=\frac{10+3D}{4}$, which leads to 
$\kappa_1=\frac{6+D}{2}$, 
$\kappa_2=\frac{18-D}{2}$, 
$\kappa_4=2$,
\item
$\kappa_4$ vanishes; $\Delta=\frac{8+3D}{4}$, which leads to 
$\kappa_1=\frac{4+D}{2}$, 
$\kappa_2=\frac{12-D}{2}$, 
$\kappa_3=-2$,
\end{enumerate}
where the value of $D$ above is $2$ since $D$ is taken to $2$ in this study. 
Therefore, when we take $\Delta$ in the case a) as
\begin{eqnarray}\label{delfx}
\Delta = (2+D)/2=2, \quad \textrm{where $D=2$ in this study.} 
\end{eqnarray}
a fixed-point exists, on which the EH action takes 
\begin{eqnarray}\label{njomzkmtr}
\lim_{n \to \infty} S^{(n)}(\zeta)    
\!\!&=&\!\!
- 
\frac{D}{16\pi G_N'}
\int d^D\zeta \sqrt{-g^{(\infty)}\left(\zeta \right)} \, 
g^{(\infty)}{}^{\mu\nu}\left(\zeta \right) \, 
\dl_\mu \phi^{(\infty)} \left(\zeta \right) \, \dl_\nu \phi^{(\infty)} \left(\zeta \right),  
\end{eqnarray}  
where we have performed the contraction: 
$
g^{(n)}{}^{\sigma\tau}\left(\zeta \right)
g^{(n)}{}^{\rho\zeta}\left(\zeta \right)
g^{(n)}_{\zeta\sigma}\left(\zeta \right)  
g^{(n)}_{\rho\tau}\left(\zeta \right) = D
$.

We can see (\ref{njomzkmtr}) is a massless free theory, thus 
the fixed-point we have obtained is a trivial one. 
Note that the existence of a trivial fixed-point is not trivial for us who will consider the Fisher metric 
as written in Sec.\ref{Chap:Intro}. Two constants are included in the EH action (\ref{RmnEHT}), 
which are Newton coupling constant  and cosmological constant. 

First,  how the cosmological constant will be is unclear.  
This is because we cannot evaluate ${\cal N}_0$ (\ref{defcalNu}).
Next, the value of the Newton coupling constant at this fixed-point should be  $G_N$, 
therefore we can see that $G_N'$ should be as follows:
\begin{eqnarray}  \label{gndegn}
G'_N = D G_N.
\end{eqnarray}  
It is considered that the value above is the value of the Newton coupling constant at the high energy limit (ultra-violet fixed point). 
Thus, we can see that the trivial fixed-point at the low energy limit (\ref{njomzkmtr}) is connected to the asymptotic free at the high energy limit. 

Here, one may consider that 
since $G_N'$ in (\ref{njomzkmtr}) can be changed to any values just by the constant rescaling of $\phi^{(\infty)} \left(\zeta \right)$, 
it is meaningless to conclude like (\ref{gndegn}). 
However it would be a misunderstanding.  
This is because we are considering the relation between (\ref{RmnEHT}) and (\ref{njomzkmtr}), 
and the problem of how to determine $G_N'$ always exists either performing the rescaling or not. 
Concretely, for example let us consider the case of the  rescaling,  
$
\phi^{(\infty)} \left(\zeta \right) 
\to   
4\sqrt{\frac{\pi G_N'}{D}}
\phi^{(\infty)} \left(\zeta \right),
$ so that the front factor of  (\ref{njomzkmtr}) becomes totally $-1$. 
In this case the front factor in (\ref{RmnEHT}) would be given by $D^{-1}$, 
and at the stage of  (\ref{njomzkmtr}) the problem of how to determine $G_N'$ exists as how to concretely determine the rescaling factor.   
If we wanted to give the front factor of (\ref{njomzkmtr}) to be $-1$,  
we would  have to determine $G_N'$  so that $4\sqrt{\frac{\pi G_N'}{D}}$ can become $1$, which leads to $G_N'=\frac{D}{16\pi}$. 
This  result is the same one in the case that we did not perform the rescaling as can be from (\ref{njomzkmtr}). 
After all, rescaling and coarse-graining would be another issues each other.

Now we have considered the case a), there exists no fixed-points in other cases b), c) and d), since some terms get diverged in these other cases.
\newline

Let us turn to $\phi^{(n)}$ and $p^{(n)}$ ($p^{(n)}$ mean the $p$ getting $n$ times coarse-grainings)  
to give $g^{(n)}{}^{\mu\nu}$ and $g^{(n)}_{\mu\nu}$ in (\ref{cogrgmn1}) and (\ref{costgmn2}). 
To this purpose, let us note the two facts: 
\begin{enumerate}
\renewcommand{\labelenumi}{\arabic{enumi}).}
\item 
Components of $g^{(n)}{}^{\mu\nu}$ and $g^{(n)}_{\mu\nu}$ are given just by exchanging $\theta^\mu$ in the components of (\ref{mt22x00}) 
with $L\zeta^\mu$ as can be seen in (\ref{cogrgmn1}) and (\ref{costgmn2}).
\item 
$\phi$ in (\ref{pAdS2FmPbt}) gives (\ref{mt22x00}). 
\end{enumerate}
Form those, 
we can reach the following $\phi^{(n)}$ to give $g^{(n)}{}^{\mu\nu}$ and $g^{(n)}_{\mu\nu}$ in (\ref{cogrgmn1}) and (\ref{costgmn2}) as 
\begin{eqnarray}\label{pAdS2FmPzt}
\phi^{(n)}(\zeta) = \frac{1}{2} \ln \left[ -\frac{\pi}{L^n \zeta^2} \right] - \frac{L^n(\zeta^1)^2}{4\zeta^2}.  
\end{eqnarray}
We can confirm that we can obtain $g^{(n)}{}^{\mu\nu}$ and $g^{(n)}_{\mu\nu}$ 
in (\ref{cogrgmn1}) and (\ref{costgmn2}) 
from $\dl/\dl \zeta^\mu \, \dl/\dl \zeta^\nu \phi^{(n)}(\zeta)$ and 
$g^{(n)}{}^{\mu\sigma}g^{(n)}{}^{\mu\rho} \dl/\dl \zeta^\sigma \, \dl/\dl \zeta^\rho \phi^{(n)}(\zeta)$
according to (\ref{metexpfm}).
\newline

Now we can see from (\ref{pAdS2FmPzt}) that $\phi^{(n)}(\zeta) \sim - \frac{L^n(\zeta^1)^2}{4\zeta^2} +{\cal O}(n\ln L)$ at $n \to \infty$. 
Therefore, with (\ref{cogrgmn1}) and (\ref{costgmn2}), we can see (\ref{njomzkmtr}) has the $L$-dependence, $L^{n(-D/2+4)}$. 
Therefore, the value of (\ref{njomzkmtr}) appears to get diverged at $n \to \infty$. 
However, (\ref{njomzkmtr})  is finite for the fact $\theta^\mu = L^n \zeta^\mu$. 
\newline

$\phi^{(n)}(\zeta)$ are linked with $\phi^{(0)}(\xi)$ by the relation (\ref{phzlphn}),  
and $\phi^{(0)}(\xi)$ is given in (\ref{pAdS2FmPbt}). 
We can see that when the forms of $\phi^{(n)}(\zeta)$ and $\phi^{(0)}(\xi)$ are given as (\ref{phzlphn}) and (\ref{pAdS2FmPbt}), 
only when $\Delta=D$, (\ref{phzlphn}) can be held; 
if the value of $\Delta$ is some other values, 
$\phi^{(0)}(\xi)$ given as (\ref{pAdS2FmPbt}) and $\phi^{(n)}(\zeta)$ given as (\ref{phzlphn}) do not satisfy the relation (\ref{phzlphn}).  
Whether or not $\Delta$ can be determined to the proper value in this sense is highly nontrivial, 
as mention the reason under (\ref{sgcgrsn}).
\newline

From the description in Sec.\ref{Sec:EPO}, $p^{(n)}$ can be also known, which is that with changing $\sigma_0^2$ with $\sigma_0^2/L^n$ as
\begin{align}\label{pAdS2Fnzt}
p^{(n)}
= \frac{1}{\sigma_0} \sqrt{\frac{L^n}{2\pi}} \, \exp \left[ -\frac{L^n}{2\sigma_0^2}(x-\bx)^2 \right].   
\end{align}
Therefore, under $n$ times coarse-grainings, $\sigma_0^2$ is considered to get the change as
\begin{align}\label{sgcgne}
\sigma_0^2 \, \to \, \sigma_0^2/L^{2n}.
\end{align}

Lastly, we can see $\bga^{(n)}$ is given by the $n$ times coarse-grainings toward $\bga^{(0)}$ as
\begin{align}\label{bgcgn}
\bga^{(n)} =- L^{-(D-\Delta)}\zeta^\mu \bF_\mu^{(n)}(x)+\phi^{(n)}(\zeta). 
\end{align}
Again, only when we consider $\Delta=D$ given in (\ref{delfx}), 
we can obtain (\ref{cogrgmn1}) and (\ref{costgmn2}) from (\ref{pAdS2Fnzt}) and (\ref{bgcgn}) according to (\ref{DefFishmet1}) for arbitrary $n$. 
Again, we can say $D$ should be determined to $2$ for this sense. It is highly nontrivial that we can obtain a conclusion that $D$ should be $2$.

%================================================================================================================== 
\section{Summary}
\label{Sec:Summary}
%================================================================================================================== 
 
We summarize this study. First of all, we have been interested to consider the gravitational theory in terms 
of some statistical averages. From this viewpoint, we have employed the Fisher metric, 
$g_{\mu\nu}(\theta) = \left< \dl_\mu \bga(x,\theta) \dl_\nu \bga(x,\theta) \right>$, 
where $\bga(x,\theta) = - \theta^\mu \bF_\mu (x) + \phi(\theta)$.

In this study, considering $\phi(\theta)$ on a space $\Lambda$, 
we have considered $\phi(\theta)$ as the underlying entity of the metrics. 
What we have done in this study are the following three:
\begin{enumerate}
\renewcommand{\labelenumi}{\arabic{enumi}).}
\item 
Rewriting the EH action in terms of the ingredients in the Fisher metric,   
\item 
determining the transformation rules of the Fisher metric, etc under the coarse-graining, 
performing the coarse-graining toward that rewritten EH action, 
\item 
based on that, obtaining a trivial fixed-point and the value of Newton constant 
at the high energy limit (ultra-violet fixed-point) so that the value of Newton constant at the trivial fixed-point is given as $G_N$.   
Here, the existence of a trivial fixed-point is not trivial for us as written in Sec.\ref{Chap:Intro}. 
\end{enumerate}

In general, performing coarse-grainings and renormalization transformations to $\phi^4$-theory effectively generate 
$\phi^i$ ($i=6,8,10,\cdots$) terms (this $\phi$ is irrelevant of that in this study). 
Whether these terms are appearing  or not in the (\ref{rehacg37}) in our study is unclear, because it is not the form given 
by $\phi^i$ ($i=6,8,10,\cdots$). Correspondingly, it is unclear that it 
can be equivalently written into the EH action with the higher order derivative corrections. (When the value of $\Delta$ is 
given as (\ref{delfx}), (\ref{rehacg37})  can reach a free massless Klein-Gordon form at the low-energy limit.)

The space-time this study has considered is two-dimensional. 
Besides, there are the following problems in this study: 
\begin{enumerate}
\renewcommand{\labelenumi}{\arabic{enumi}).}
%-----
\item 
The action we have rewritten in terms of the Fisher metric 
has not been the form that we can rewrite into the momentum 
space using the Fourier transformation,  
%-----
\item 
the quadratic part giving the two-point correlated function did not exist in the action 
if we look at the action in terms of $\phi(\theta)$,  
%-----
\item 
cosmological constant at the fixed-point is unclear since ${\cal N}_0$ in (\ref{defcalNu}) cannot be evaluated\footnote{ 
Corresponding to this, whether the cosmological constant can 
be appropriately generated when the AdS metric is formed so that the AdS metric can be a solution would be a question, 
where it is shown that the Fisher metric obtained from $p$ in (\ref{pAdS2Fmbt}) can form a two-dimensional the AdS metric 
in the Poincar$\grave{\rm e}$ coordinate under the condition $\frac{(\theta^1)^2}{\theta^2} \ll 1$ 
\cite{Matsueda:2013saa,Matsueda:2014lda,Matsueda:2015voa}.}   
%-----
\item 
the problem of the number of the degree of freedom of the fisher metric, which we mention at the last of Sec.\ref{Sec:What}.
%-----

%-----
\item  
if we consider quantum fluctuation on $\phi$, the fundamental constraint (\ref{sump1}) gets broken 
(see Sec.\ref{Sec:Renomalization} for more detail). 
\end{enumerate}

1) and 2) are important in performing the renormalization transformation, 
and for these reason, we have sufficed it to perform the coarse-graining 
in this study. 

An interesting direction led from this study is to consider some correspondence between the 3D and 2D AdS gravitational theories. 
This is because  3D AdS garvitational theory and 2D CFT is linked by the Kerr/CFT correspondence \cite{Guica:2008mu}, and 2D 
CFT and 2D gravitational theory is linked by this study.
\newline

Lastly, we would like to give a comment about the relation with the \cite{Verlinde:2010hp} we refer in the introduction. 

We have obtained the description of the two-dimensional gravitational theory 
at the low energy limit by a massless free field theory  (as for ``the low energy limit'', see under (\ref{CGO})). 
By this, as one of things we can say, it follows that we have 
obtained the description of the gravitational theory on the two-dimensional branes forming the foliation structure 
of the space-time at the low-energy limit.

In the \cite{Verlinde:2010hp}, the gravitational force in the perpendicular direction toward the branes are explained 
as the entropic force arisen from the difference of the entropy on the surface of the adjacent branes, however there 
has been no explanation for the gravitational force working transversely on the brane. 
In such a situation, the description by the free scalar field we have obtained would be meaningful.

It is interesting as the future direction  to check whether or not the gravitational force described as the entropic force can 
be calculated by the entropy of the scalar field in this study. In that study, we would first calculate the entropy of the free 
massless scalar field on the branes.

\paragraph*{Acknowledgment ---} \hspace{-3mm}
Although I have pointed out some problems in \cite{Matsueda:2013saa} in the last of the introduction,  
it (and \cite{Matsueda:2014lda}, \cite{Matsueda:2015voa}) has been technically a big help in this study.

\appendix 
 
%================================================================================================================== 
\section{Expression of the Ricci tensor}
\label{App:CalRic}
%================================================================================================================== 
 
In Appendix.\ref{App:CalEin}, 
we derive the expression of the Ricci tensors when $p=e^{-\bga}$ as in (\ref{defpth}). 
The final result is (\ref{ExpRicibga}). 
Based on that, in Appendix.\ref{App:CalEinbga}, 
we obtain the expression of the Ricci tensor in (\ref{Xmntr3}) 
when $\bga$ is given as $\bga = -\theta^\mu \bF_\mu + \phi$ as in (\ref{dforggm}). 
(A large part of the description in this Appendix is overlapped with \cite{Matsueda:2013saa}.)

%================================================================================================================== 
\subsection{Expression of the Ricci tensor when $p=e^{-\bga}$}
\label{App:CalEin}
%================================================================================================================== 

We first obtain the expression of the Christoffel symbols: 
\begin{align}\label{OurCrs}
\Gamma^\lambda_{\mu\nu} 
= 
\frac{1}{2} \, g^{\lambda\tau} 
\left( \dl_\mu g_{\nu \tau} + \dl_\nu g_{\mu \tau} - \dl_\tau g_{\mu \nu} \right)
\end{align} 
in terms of $\bga$. 
Here, as mentioned in Sec.\ref{Sec:What}, 
we write $\gamma$ and $g_{\mu\nu}$ 
before taking the statistical average in boldface.  
Then, from (\ref{DefFishmet1}), we can see 
\begin{align}\label{gchrisbas}
\dl_{\sigma} g_{\mu\nu} 
= 
- \left< \dl_{\sigma} \bga \dl_{\mu}\bga \dl_{\nu} \bga \right> 
+ \left< (\dl_{\sigma} \dl_{\mu}\bga)    \, \dl_{\nu} \bga \right> 
+ \left< \dl_{\mu} \bga                  \, (\dl_{\sigma} \dl_{\nu} \bga) \right>. 
\end{align}
Therefore,
\begin{align}\label{ctsgama} 
\Gamma^\lambda_{\mu\nu} = g^{\lambda \tau} 
\left(
\left< (\dl_\mu \dl_\nu \bga) \dl_\tau \bga \right> - \frac{1}{2} \left< \dl_\mu \bga \dl_\nu \bga \dl_\tau \bga \right>
\right). 
\end{align}

Next, let us obtain the expression of the Ricci tensors: 
\begin{align}\label{DefRicciTes}
R_{\mu\nu}
= \dl_\sigma \Gamma^\sigma_{\mu\nu}
-\dl_\nu \Gamma^\sigma_{\mu\sigma}
+\Gamma^\sigma_{\rho\sigma}\Gamma^\rho_{\mu\nu}
-\Gamma^\sigma_{\rho\nu}\Gamma^\rho_{\mu\sigma} 
\end{align}
in terms of $\bga$. 
We first write $R_{\mu\nu}$ as
\begin{align}\label{RcTgama}
R_{\mu\nu} = A_{\mu\nu} + B_{\mu\nu} + C_{\mu\nu}
\end{align}
where 
\begin{align}\label{R3ptA}
A_{\mu\nu} 
=& 
g^{\sigma\tau}\dl_{\sigma}
\left(
\left< (\dl_{\mu}\dl_{\nu}\bga)\dl_{\tau}\bga \right> - \frac{1}{2} \left< \dl_{\mu}\bga\dl_{\nu}\bga\dl_{\tau}\bga \right>
\right) 
\nonumber \\
& 
- g^{\sigma\tau}\dl_{\nu}
\left( 
\left< (\dl_{\mu}\dl_{\sigma}\bga)\dl_{\tau}\bga \right> - \frac{1}{2} \left< \dl_{\mu}\bga\dl_{\sigma}\bga\dl_{\tau}\bga \right>
\right), 
\end{align}
\begin{align}\label{R3ptB}
B_{\mu\nu} 
=& \,\,\,\,\,\,
\left(\dl_{\sigma}g^{\sigma\tau}\right)
\left(
\left< (\dl_{\mu}\dl_{\nu}\bga)\dl_{\tau}\bga \right> - \frac{1}{2} \left< \dl_{\mu}\bga\dl_{\nu}\bga\dl_{\tau}\bga \right>
\right) 
\nonumber \\
& 
- \left(\dl_{\nu}g^{\sigma\tau}\right)
\left(
\left< (\dl_{\mu}\dl_{\sigma}\bga)\dl_{\tau}\bga \right> - \frac{1}{2} \left< \dl_{\mu}\bga\dl_{\sigma}\bga\dl_{\tau}\bga \right>
\right),
\end{align}
\begin{align}\label{R3ptC}
C_{\mu\nu} 
=& \,\,\,\,\,\,
g^{\sigma\tau}g^{\rho\zeta}
\left(
\left< (\dl_{\rho}\dl_{\sigma}\bga)\dl_{\tau}\bga \right> - \frac{1}{2} \left< \dl_{\rho}\bga\dl_{\sigma}\bga\dl_{\tau}\bga \right>
\right) 
\left(
\left< (\dl_{\mu}\dl_{\nu}\bga)\dl_{\zeta}\bga \right> - \frac{1}{2} \left< \dl_{\mu}\bga\dl_{\nu}\bga\dl_{\zeta}\bga \right> 
\right) 
\nonumber \\
&  
- g^{\sigma\tau}g^{\rho\zeta}
\left(
\left< (\dl_{\rho}\dl_{\nu}\bga)\dl_{\tau}\bga \right> -\frac{1}{2} \left< \dl_{\rho}\bga\dl_{\nu}\bga\dl_{\tau}\bga \right>
\right) 
\left(
\left< (\dl_{\mu}\dl_{\sigma}\bga)\dl_{\zeta}\bga \right> - \frac{1}{2} \left< \dl_{\mu}\bga\dl_{\sigma}\bga\dl_{\zeta}\bga \right>
\right).
\end{align}
We can rewrite $A_{\mu\nu}$ in (\ref{R3ptA}) as  
\begin{align}
\label{R3ptAp}
A_{\mu\nu} 
=& \,\,
g^{\sigma\tau}
\left\{
- \, \frac{1}{2} 
\left< \dl_{\sigma}\bga(\dl_{\mu}\dl_{\nu}\bga)\dl_{\tau}\bga \right> 
+ \left< (\dl_{\mu}\dl_{\nu}\bga)(\dl_{\sigma}\dl_{\tau}\bga) \right> 
\right.\nonumber \\
%----------
& \,\,\,\,\,\,\,\,\,\,\,\,\,\,
+ \frac{1}{2}
\left< \dl_{\nu}\bga (\dl_{\mu}\dl_{\sigma}\bga)\dl_{\tau}\bga \right> - 
\left< (\dl_{\mu}\dl_{\sigma}\bga)(\dl_{\nu}\dl_{\tau}\bga) \right> 
\nonumber \\
%----------
& \,\,\,\,\,\,\,\,\,\,\,\,\,\,
\left. 
- \, \frac{1}{2} \left< \dl_{\mu}\bga \dl_{\nu}\bga (\dl_{\sigma}\dl_{\tau}\bga) \right> 
+ \frac{1}{2} \left< \dl_{\mu}\bga \dl_{\sigma}\bga (\dl_{\nu}\dl_{\tau}\bga) \right>
\right\}
\nonumber \\
%==========
=& \,\,\,\,
\left< \bomg \, (\dl_{\mu}\dl_{\nu}\bga) \right> 
- \frac{1}{2}g^{\sigma\tau} \left< (\dl_{\sigma}\dl_{\tau}\bga) \dl_{\mu}\bga \dl_{\nu}\bga \right> 
- g^{\sigma\tau}
\left< (\dl_{\mu}\dl_{\sigma}\bga)(\dl_{\nu}\dl_{\tau}\bga) \right>  
\nn
%---------- 
&
+ \frac{1}{2}g^{\sigma\tau}
\left< 
\dl_{\sigma}\bga\dl_{\mu}\bga (\dl_{\nu}\dl_{\tau}\bga) + \dl_{\sigma}\bga\dl_{\nu}\bga (\dl_{\mu}\dl_{\tau}\bga)
\right>, 
\end{align}
where
\begin{align}\label{bomg}
\bomg
\equiv
g^{\sigma\tau} \left( \dl_{\sigma}\dl_{\tau}\bga-\frac{1}{2}\dl_{\sigma}\bga \dl_{\tau}\bga \right). 
\end{align}
%\newpage
We here would like to note two points in the footnote
\footnote{
%****************************************
%*************** FOOTNOTE ***************
%****************************************
\begin{itemize}
\item
We cannot deform $\bomg$ above to $\frac{1}{2}g^{\sigma\tau}\bg_{\sigma\tau}$, 
because $\dl_{\sigma}\dl_{\tau}\bga$ and $\dl_{\sigma}\bga \dl_{\tau}\bga$ are different 
before taken the statistical average (\ref{ladddrlstav}).  
\item
Next, even if we could deform to $\frac{1}{2}g^{\sigma\tau}\bg_{\sigma\tau}$,
we could not deform as $g^{\sigma\tau}\bg_{\sigma\tau}=n$ by performing the contraction. 

The metrices before taken the statistical average which we denote in the bold face as $\bg^{\mu\nu}(x, \theta)$ 
would be always metrics of some spaces whatever $x$. 
%-----
However, $\bg_{\mu\nu}(x, \theta)$ and $g_{\mu\nu}(\theta)$ are associated with different spaces as the metric; 
$\bg_{\mu\nu}(x, \theta)$ are one of some possible metrices and $g_{\mu\nu}(\theta)$ are 
the metrices of the space appearing after taken the statistical average. 
%-----
Therefore, $\bg_{\mu\nu}(x, \theta)$ and $g_{\mu\nu}(\theta)$ are not in the relation of the inverse matrix each other. 
\end{itemize}
%****************************************
%****************************************
%****************************************
}.

We can also rewrite $B_{\mu\nu}$ in (\ref{R3ptB}) as  
\begin{align}
B_{\mu\nu} 
=& 
- g^{\sigma\tau}g^{\rho\zeta}
\left(
- \left< \dl_{\sigma}\bga\dl_{\tau}\bga \dl_{\rho}\bga \right>
+ \left< (\dl_{\sigma}\dl_{\tau}\bga) \dl_{\rho}\bga \right> + \left< \dl_{\tau}\bga (\dl_{\sigma}\dl_{\rho}\bga) \right>
\right) 
\nonumber \\
& 
\hspace{10.5mm}
\times
\left(
\left< (\dl_{\mu}\dl_{\nu}\bga)\dl_{\zeta}\bga \right> - \frac{1}{2} \left< \dl_{\mu}\bga\dl_{\nu}\bga\dl_{\zeta}\bga \right>
\right) 
\nonumber \\
& 
+ g^{\sigma\tau}g^{\rho\zeta}
\left(
- \left< \dl_{\nu}\bga\dl_{\tau}\bga\dl_{\rho}\bga \right>
+ \left< (\dl_{\nu}\dl_{\tau}\bga)\dl_{\rho}\bga \right>
+ \left<\dl_{\tau}\bga(\dl_{\nu}\dl_{\rho}\bga)
\right>
\right) \nonumber \\
& 
\hspace{10.5mm} 
\times
\left(
\left< (\dl_{\mu}\dl_{\sigma}\bga)\dl_{\zeta}\bga \right> - \frac{1}{2} \left< \dl_{\mu}\bga \dl_{\sigma}\bga \dl_{\zeta}\bga \right>
\right),
\end{align}
where we have used a general relation $\dl_{\sigma}g^{\mu\nu}=-g^{\mu\alpha}g^{\nu\beta}\dl_{\sigma}g_{\alpha\beta}$. 
Summing up $B_{\mu\nu}$ above with $C_{\mu\nu}$ in (\ref{R3ptC}),
\begin{align}\label{R3ptBC}
& 
B_{\mu\nu}+C_{\mu\nu} \nn
=& \,\,\,\,\,\,\,
g^{\sigma\tau}g^{\rho\zeta}
\left(
\left< (\dl_{\mu}\dl_{\sigma}\bga)\dl_{\zeta}\bga \right> - \frac{1}{2} \left< \dl_{\mu}\bga \dl_{\sigma}\bga \dl_{\zeta}\bga  \right>
\right) 
\left(
\left< (\dl_{\nu}\dl_{\tau}\bga)\dl_{\rho}\bga \right>
-\frac{1}{2}
\left< \dl_{\nu}\bga \dl_{\tau}\bga \dl_{\rho}\bga \right>
\right) 
\nonumber \\
& 
- g^{\sigma\tau}g^{\rho\zeta}
\left(\left< (\dl_{\sigma}\dl_{\tau}\bga)\dl_{\rho}\bga \right>
-\frac{1}{2}
\left< \dl_{\sigma}\bga \dl_{\tau}\bga \dl_{\rho}\bga \right>
\right) 
\left(
\left< (\dl_{\mu}\dl_{\nu}\bga)\dl_{\zeta}\bga \right>
-\frac{1}{2}
\left< \dl_{\mu}\bga \dl_{\nu}\bga \dl_{\zeta}\bga \right>
\right) 
\nonumber \\ 
=& \,\,\,\,\,\,\, 
g^{\sigma\tau}g^{\rho\zeta}
\left< \dl_{\zeta}\bga \left( \dl_{\mu} \dl_{\sigma} \bga - \frac{1}{2} \dl_{\mu} \bga \dl_{\sigma}\bga \right) \right>
\left< 
\dl_{\rho}\bga \left( \dl_{\nu}\dl_{\tau}\bga-\frac{1}{2} \dl_{\nu}\bga \dl_{\tau}\bga \right)
\right>
\nonumber \\
& 
- g^{\sigma\tau}g^{\rho\zeta}
\left<
\dl_{\rho}\bga
\left( \dl_{\sigma}\dl_{\tau}\bga - \frac{1}{2} \dl_{\sigma}\bga \dl_{\tau}\bga \right)
\right>
\left<
\dl_{\zeta}\bga
\left( \dl_{\mu}\dl_{\nu}\bga - \frac{1}{2} \dl_{\mu}\bga \dl_{\nu}\bga \right)
\right>
\nonumber \\ 
=& \,\,\,\,\,\,\, 
g^{\sigma\tau}g^{\rho\zeta}
\left< \dl_{\zeta}\bga \left( \dl_{\mu} \dl_{\sigma} \bga - \frac{1}{2} \dl_{\mu} \bga \dl_{\sigma}\bga \right) \right>
\left< \dl_{\rho}\bga \left( \dl_{\nu}\dl_{\tau}\bga-\frac{1}{2} \dl_{\nu}\bga \dl_{\tau}\bga \right) \right>
\nonumber \\
& 
+ 
\left< \bp \left( \dl_{\mu}\dl_{\nu}\bga - \frac{1}{2} \dl_{\mu}\bga \dl_{\nu}\bga \right) \right>,
\end{align}
where 
\begin{align}\label{bp}
\bp  
\equiv  
-g^{\sigma\tau}g^{\rho\zeta}\dl_{\zeta}\bga
\left< \dl_{\rho}\bga \left( \dl_{\sigma}\dl_{\tau}\bga-\frac{1}{2}\dl_\sigma \bga \dl_\tau \bga \right) \right>.
\end{align}

From (\ref{R3ptAp}) and (\ref{R3ptBC}), we can obtain the expression of the Ricci tensor in $p=e^{-\bga}$ as  
\begin{align}\label{ExpRicibga}
R_{\mu\nu} 
=& \,\,\,\,\,\,
\left< \left(\bomg+\bp\right) \dl_{\mu}\dl_{\nu}\bga \right> 
- \frac{1}{2}
\left< \left( g^{\sigma\tau}(\dl_{\sigma}\dl_{\tau}\bga) + \bp \right) \dl_{\mu} \bga \dl_{\nu} \bga \right>  
- g^{\sigma\tau}
\left< (\dl_{\mu}\dl_{\sigma}\bga) (\dl_{\nu}\dl_{\tau}\bga) \right> 
\nonumber \\
& 
+ \frac{1}{2}g^{\sigma\tau}
\left< 
\dl_{\sigma} \bga \dl_{\mu} \bga (\dl_{\nu}\dl_{\tau}\bga) + \dl \bga_{\sigma} \dl \bga_{\nu} (\dl_{\mu}\dl_{\tau}\bga)
\right> \nonumber \\
& 
+ g^{\sigma\tau}g^{\rho\zeta}
\left< \dl_{\zeta}\bga \left( \dl_{\mu}\dl_{\sigma}\bga-\frac{1}{2} \dl_{\mu} \bga \dl_{\sigma} \bga \right) \right>
\left< \dl_{\rho}\bga \left( \dl_{\nu}\dl_{\tau}\bga-\frac{1}{2} \dl_{\nu} \bga \dl_{\tau} \bga \right) \right>. 
\end{align}

%================================================================================================================== 
\subsection{Expression of the Ricci tensor when $\bga = -\theta^\mu \bF_\mu + \phi$}
\label{App:CalEinbga}
%================================================================================================================== 

In this appendix, we obtain the expression of the Ricci tensor 
when $\bga = -\theta^\mu \bF_\mu + \phi$ as in (\ref{dforgp}) based on (\ref{ExpRicibga}). 
The final result is (\ref{nzlsmt}), which leads to (\ref{Xmntr3}).
We exploit the relations in Sec.\ref{Sec:EPG}. 
\newline

Upon evaluating (\ref{ExpRicibga}), we first calculate the statistical average of $\bomg$ and $\bp$ 
in (\ref{bomg}) and (\ref{bp}) as
\begin{eqnarray}
\label{bomgav}
\left< \bomg \right> \! &=& \! g^{\sigma\tau}
\left(\dl_\sigma \dl_\tau \phi - \frac{1}{2} \left< \dl_\sigma \bga \dl_\tau \bga \right>\right) =\frac{D}{2}, \\
%---
\label{bomgzr}
\left< \bp \right> \! &=& \! 0. 
\end{eqnarray}
We also calculate the term in the last line in (\ref{ExpRicibga}) as
\begin{eqnarray}\label{ricllap}
&& 
\left< \dl_{\zeta}\bga \left( \dl_\mu \dl_\sigma \bga-\frac{1}{2} \dl_\mu \bga \dl_\sigma \bga \right) \right>
\left< \dl_\rho \bga \left( \dl_\nu \dl_\tau \bga    -\frac{1}{2} \dl_\nu \bga \dl_\tau \bga \right) \right>
\nn
\! &=& \!
\left( \left< \dl_\zeta \bga \right>  (\dl_\mu \dl_\sigma \phi) -\frac{1}{2} \left< \dl_\zeta \bga \dl_\mu \bga \dl_\sigma \bga  \right> \right)
\left( \left< \dl_\sigma \bga \right> (\dl_\nu \dl_\tau \phi)   -\frac{1}{2} \left< \dl_\rho \bga \dl_\nu \bga \dl_\tau \bga  \right>  \right)
\nn
\! &=& \!
\frac{1}{4} \left< \dl_\zeta \bga \dl_\mu \bga \dl_\sigma \bga  \right>\left< \dl_\rho \bga \dl_\nu \bga \dl_\tau \bga  \right>. 
\end{eqnarray}
Then, 
\begin{eqnarray}\label{riclap}
{\rm (\ref{ExpRicibga})}
&=& \!
\big< \left( \bomg + \bp \right) \dl_\mu \dl_\nu \bga \big> 
- 
\frac{1}{2}
\big< 
\left( g^{\sigma\tau} \dl_\sigma \dl_\tau \bga + \bp \right) 
\dl_\mu \bga \dl_\nu \bga 
\big>  
- 
g^{\sigma\tau}
\big< (\dl_\mu \dl_\sigma \bga) (\dl_\nu \dl_\tau \bga) \big> 
\nn 
%----------
&&+ \,\,
\frac{1}{2}
g^{\sigma\tau}
\big< 
\dl \bga_{\sigma} \dl \bga_\mu (\dl_\nu \dl_\tau \bga) + \dl_\sigma \bga \dl_\nu \bga (\dl_\mu \dl_\tau \bga) 
\big> 
\nn 
%----------
&&+ \,\,
\frac{1}{4} g^{\sigma\tau} g^{\rho\xi} 
\left< \dl_\zeta \bga \dl_\mu \bga \dl_\sigma \bga  \right>
\left< \dl_\rho \bga \dl_\nu \bga \dl_\tau \bga  \right>
\nn
%==========
&=& \!
\frac{D}{2} \, \dl_\mu \dl_\nu \bga
- \frac{1}{2}
\big( 
g^{\sigma\tau} \dl_\sigma \dl_\tau \bga \left< \dl_\mu \bga \dl_\nu \bga \right> 
+
\left< \bp \dl_\mu \bga \dl_\nu \bga \right> 
\big) 
\nn 
%----------
&& +\,\,
\frac{1}{4} g^{\sigma\tau} g^{\rho\xi} 
\left< \dl_\zeta \bga \dl_\mu \bga \dl_\sigma \bga  \right>\left< \dl_\rho \bga \dl_\nu \bga \dl_\tau \bga \right>,
\end{eqnarray}
where we have used the fact that $\dl_\mu \dl_\nu \bga$ is independent of $x$ as in (\ref{ggdrl2}) 
when $\bga = -\theta^\mu \bF_\mu + \phi$ (we also proceed with the following calculation using this relation), 
and 
\begin{eqnarray}
\label{riclapud1}
g^{\sigma\tau}
\left< (\dl_\mu \dl_\sigma \bga) (\dl_\nu \dl_\tau \bga) \right> &=& g_{\mu\nu},\\
\label{riclapud2}
g^{\sigma\tau}
\left< \dl \bga_{\sigma} \dl \bga_\mu (\dl_\nu \dl_\tau \bga) + \dl_\sigma \bga \dl_\nu \bga (\dl_\mu \dl_\tau \bga) \right> &=& 2g_{\mu\nu}. 
\end{eqnarray}
Continuing the evaluation, 
\begin{eqnarray}\label{rdbgdbgdbgah}
{\rm (\ref{riclap})}   
\!\! &=& \!\!
\frac{D}{2} \dl_\mu \dl_\nu \bga 
- \frac{1}{2}
\left( n \, g_{\mu\nu} 
+ \left< \bp \dl_\mu \bga \dl_\nu \bga \right> \right) 
+\frac{1}{4} g^{\sigma\tau} g^{\rho\zeta} 
\left< \dl_\zeta \bga \dl_\mu \bga \dl_\sigma \bga  \right>
\left< \dl_\rho \bga \dl_\nu \bga \dl_\tau \bga  \right> \nn
%==========
\!\! &=& \!\! 
-\frac{1}{4} g^{\sigma\tau} g^{\rho\zeta}  
\left< \dl_\rho \bga \dl_\sigma \bga \dl_\tau \bga \right>
\left< \dl_\zeta \bga \dl_\mu \bga \dl_\nu \bga \right>
+\frac{1}{4} g^{\sigma\tau} g^{\rho\zeta}  
\left< \dl_\zeta \bga \dl_\mu \bga \dl_\sigma \bga \right>
\left< \dl_\rho \bga \dl_\nu \bga \dl_\tau \bga  \right>, \nn
\end{eqnarray}
where, when $\bga = -\theta^\mu \bF_\mu + \phi$, $\bp$ can be evaluated as
\begin{eqnarray}
\bp &=& -g^{\sigma\tau}g^{\rho\zeta}\dl_\zeta \bga 
\left( 
\left< \dl_\rho (\dl_\sigma \dl_\tau \bga) \right>
-\frac{1}{2} \left< \dl_\rho \bga \dl_\sigma \bga \dl_\tau \bga \right>
\right)
\nn 
%----------
&=& 
\frac{1}{2} g^{\sigma\tau}g^{\rho\zeta}\dl_\zeta \bga \left< \dl_\rho \bga \dl_\sigma \bga \dl_\tau \bga \right>
\end{eqnarray}
and then, 
\begin{eqnarray}
\left< \bp \, \dl_\mu \bga \dl_\nu \bga \right>
=
\frac{1}{2} \, g^{\sigma\tau} g^{\rho\xi} 
\left< \dl_\rho \bga \dl_\sigma \bga \dl_\tau \bga \right>
\left< \dl_\zeta \bga \dl_\mu \bga \dl_\nu \bga \right>. 
\end{eqnarray}

Evaluating each term in (\ref{rdbgdbgdbgah}) as
\begin{eqnarray}
&&
g^{\sigma\tau} g^{\rho\zeta}  
\left< \dl_\rho \bga \dl_\sigma \bga \dl_\tau \bga \right>
\left< \dl_\zeta \bga \dl_\mu \bga \dl_\nu \bga  \right> 
\nn
%=====
[1.5mm]
&=& 
g^{\sigma\tau} g^{\rho\zeta}  
\left< \dl_\rho \bga \dl_\sigma \bga \dl_\tau \bga \right>
\left< \dl_\zeta \bga \left(\dl_\mu \phi-\bF_\mu\right) \left(\dl_\nu \phi-\bF_\nu\right) \right> 
\nn
%=====
[1.5mm]
&=&
g^{\sigma\tau} g^{\rho\zeta}  
\left< \dl_\rho \bga \dl_\sigma \bga \dl_\tau \bga \right>
\big( 
- \left< \dl_\zeta \bga \bF_\mu \right> \dl_\nu \phi 
- \left< \dl_\zeta \bga \bF_\nu \right> \dl_\mu \phi
+ \left< \dl_\zeta \bga\bF_\mu \bF_\nu \right>
\big), 
\\
%********************
[5.0mm] 
&&
g^{\sigma\tau} g^{\rho\zeta}  
\left< \dl_\zeta \bga \dl_\mu \bga \dl_\sigma \bga \right>
\left< \dl_\rho \bga \dl_\nu \bga \dl_\tau \bga \right> 
\nn
%=====
[1.5mm]
&=& 
g^{\sigma\tau} g^{\rho\zeta}  
\left< \dl_\zeta \bga \left(\dl_\mu \phi-\bF_\mu\right) \dl_\sigma \bga \right>
\left< \dl_\rho \bga  \left(\dl_\nu \phi-\bF_\nu\right) \dl_\tau \bga \right> 
\nn
%=====
[1.5mm]
&=& 
g^{\sigma\tau} g^{\rho\zeta}  
\left( g_{\zeta\sigma} \dl_\mu \phi - \left<\dl_\zeta \bga \dl_\sigma \bga \bF_\mu \right> \right)
\left( g_{\rho\tau}    \dl_\nu \phi - \left<\dl_\rho \bga \dl_\tau \bga \bF_\nu    \right> \right)  
\nn
%=====
[1.5mm]
&=& 
n \dl_\mu \phi \dl_\nu \phi 
- g^{\rho\tau} \dl_\mu \phi \left<\dl_\rho \bga \dl_\tau \bga \bF_\nu \right>
- g^{\sigma\zeta} \left<\dl_\zeta \bga \dl_\sigma \bga \bF_\mu \right> \dl_\nu \phi 
\nn
%-----
[1.5mm] 
&&+\,\,
g^{\sigma\tau}g^{\sigma\zeta} 
\left<\dl_\zeta \bga \dl_\sigma \bga \bF_\mu \right>\left<\dl_\rho \bga \dl_\tau \bga \bF_\nu \right>, 
\end{eqnarray}
we can write as
\begin{eqnarray}\label{nz21afrain}
{\rm (\ref{rdbgdbgdbgah})} 
&=&
\frac{D}{4} \, \dl_\mu \phi \dl_\nu \phi 
\nn
%-----
&&+\,\,
\frac{1}{4} 
\Big\{
- g^{\rho\tau} \dl_\mu \phi \left<\dl_\rho \bga \dl_\tau \bga \bF_\nu \right>
- g^{\sigma\zeta} \left<\dl_\zeta \bga \dl_\sigma \bga \bF_\mu \right> \dl_\nu \phi
\nn 
%-----
&&+\,\,
g^{\sigma\tau}g^{\rho\zeta} 
\left< \dl_\rho \bga \dl_\sigma \bga \dl_\tau \bga \right>
\big(
  \left< \dl_\zeta \bga \bF_\mu \right> \dl_\nu \phi
+ \left< \dl_\zeta \bga \bF_\nu \right> \dl_\mu \phi
\big)
\Big\}
\nn
%-----
&&+\,\,
\frac{1}{4} \, g^{\sigma\tau}g^{\rho\zeta} 
\Big(
\left<
\dl_\zeta \bga \dl_\sigma \bga \bF_\mu \right>\left<\dl_\rho \bga \dl_\tau \bga \bF_\nu \right>
-\left< \dl_\rho \bga \dl_\sigma \bga \dl_\tau \bga \right> \left< \dl_\zeta \bga\bF_\mu \bF_\nu \right>
\Big). \nn
\end{eqnarray}
We can calculate the terms appearing in (\ref{nz21afrain}) as
\begin{eqnarray}
\label{nzml2a}
&&
- g^{\rho\tau} \dl_\mu \phi \left<\dl_\rho \bga \dl_\tau \bga \bF_\nu \right>
- g^{\sigma\zeta} \left<\dl_\zeta \bga \dl_\sigma \bga \bF_\mu \right> \dl_\nu \phi
\nn
%-----
[1.5mm] 
&=&
- g^{\rho\tau} \dl_\mu \phi \left<\dl_\rho \bga \dl_\tau \bga (\dl_\nu \phi- \dl_\nu \bga) \right>
- g^{\sigma\zeta} \left<\dl_\zeta \bga \dl_\sigma \bga          (\dl_\mu \phi- \dl_\mu \bga) \right> \dl_\nu \phi
\nn
%-----
[1.5mm] 
&=&
- 2D \dl_\mu \phi \dl_\nu \phi 
+ g^{\rho\tau} \dl_\mu \phi 
\left<
\dl_\rho \bga \dl_\tau \bga \dl_\nu \bga
\right>
+g^{\sigma\zeta} \dl_\nu \phi 
\left<
\dl_\zeta \bga \dl_\sigma \bga \dl_\mu \bga
\right>, \\
%********************
[5.0mm] 
\label{nzml2b}
&&
g^{\sigma\tau}g^{\rho\zeta} 
\left< \dl_\rho \bga \dl_\sigma \bga \dl_\tau \bga \right>
\left(
  \left< \dl_\zeta \bga \bF_\mu \right> \dl_\nu \phi
+ \left< \dl_\zeta \bga \bF_\nu \right> \dl_\mu \phi
\right)
\nn 
%-----
[1.5mm]   
&=&
-g^{\sigma\tau}
\left( 
\left< \dl_\mu \bga \dl_\sigma \bga \dl_\tau \bga \right> \dl_\nu \phi
+
\left< \dl_\nu \bga \dl_\sigma \bga \dl_\tau \bga \right> \dl_\mu \phi
\right),
\end{eqnarray}
where we have used (\ref{exsgppFF}) in (\ref{nzml2b}).  
With those above, we can write as
\begin{eqnarray}\label{nzlsmt}
{\rm (\ref{nz21afrain})} 
&=&
\frac{D}{4} \, \dl_\mu \phi \dl_\nu \phi 
+ \frac{1}{4}\left\{ {\rm (\ref{nzml2a})} + {\rm (\ref{nzml2b})} \right\} 
+ \frac{1}{4} \, g^{\sigma\tau}g^{\rho\zeta} ( \cdots ) \nn
&=& - \frac{D}{4} \, \dl_\mu \phi \dl_\nu \phi + \frac{1}{4} g^{\sigma\tau}g^{\rho\zeta} \, ( \cdots ). 
\end{eqnarray}
Using (\ref{Nmpnp}) and so on, the one above can reach (\ref{Xmntr3}).

%================================================================================================================== 
\section{Coarse-graining in terms of $\mathfrak{g}^{(n)}_{\mu\nu}\left(\zeta \right)$ and $\mathfrak{g}^{(n)}{}^{\mu\nu}\left(\zeta \right)$}
\label{App:cgrmtg} 
%================================================================================================================== 
 
In this appendix, considering the following replacements, 
\begin{eqnarray}\label{gcgrsnac}
\mathfrak{g}^{(0)}{}^{\mu\nu}(\theta) \!\! &\to& \!\! L^{2n}\, \mathfrak{g}^{(1)}{}^{\mu\nu}(\eta), \quad 
\mathfrak{g}^{(0)}_{\mu\nu}(\theta) \,\to\, L^{-n}\, \mathfrak{g}^{(1)}_{\mu\nu}(\eta),
\end{eqnarray}  
instead of (\ref{gcgrsn}) (the grounds of this is (\ref{cogrgmn1}) and (\ref{costgmn2})), 
we show that in the case of (\ref{gcgrsnac}), we cannot obtain the proper $\Delta$, 
which means we cannot obtain the consistent results: 
the relations of (\ref{DefFishmet1}) and (\ref{metexpfm}) cannot be held 
for $\phi^{(n)}(\zeta)$ and $g^{(n)}{}^{\mu\nu}(\zeta)$ and $ g^{(n)}_{\mu\nu}(\zeta)$ obtained by performing the coarse-graining.

In this section, we do not include the replacement (\ref{sgcgrsn}), 
since determining the transformation rule of $\sigma_0^2$ would not make sense as long as $\Delta$ can be determined rightly, 
and the transformation rule of $\sigma_0^2$ is not important in the purpose in this appendix.
\newline

By the replacements with but (\ref{gcgrsn}) not (\ref{gcgrsnac}), we can obtain as
\begin{eqnarray}\label{rehacgac}
S^{(n)}(\zeta)  
= 
L^{\frac{nD}{2}} \int d^D\zeta \sqrt{-\mathfrak{g}^{(n)}\left( \zeta \right)} \, {\cal L}^{(n)}\left(\zeta \right).  
\end{eqnarray}  
where $S^{(n)}$ is the theory on $\Lambda^{(n)}$ 
with the coordinate $\zeta^\mu = \theta^\mu / L^n$ for any $n$ as defined under (\ref{cogrgmn1}).
$L^{{nD}/{2}}$ comes from $ d \theta^D \sqrt{-g(\theta)}$, 
and  ${\cal L}^{(n)} \left(\zeta \right)$ is given as  
%******************************************
\allowdisplaybreaks
%****************************************** 
\begin{eqnarray}\label{rehacg37ac}
{\cal L}^{(n)}(\zeta) 
&=&
\hspace{4mm}
L^{6n} \, 
\mathfrak{g}^{(n)}{}^{\sigma\tau}\left(\zeta \right)
\mathfrak{g}^{(n)}{}^{\rho\zeta}\left(\zeta \right)
\mathfrak{g}^{(n)}{}^{\mu\nu}\left(\zeta \right) \,  
\bigg\{ \nn
%----------
[1.5mm] 
&&
- \,
L^{-2n(D-\Delta+2)}
\mathfrak{g}^{(n)}_{\zeta\sigma}\left(\zeta \right) \, 
\mathfrak{g}^{(n)}_{\rho\tau}\left(\zeta \right) \, 
\dl_\mu \phi^{(n)}\left(\zeta \right)\, \dl_\nu \phi^{(n)}\left(\zeta \right) \nn
%----------
[1.5mm] 
&&
+\,
L^{-6n(D-\Delta+1)} \,
{\cal F}^{(n)}_{0,\mu\zeta\sigma}
\Big( {\cal P}^{(n)}_{\nu\rho} \dl_\tau \phi^{(n)}\left(\zeta \right)+ {\cal P}^{(n)}_{\nu\tau} \dl_\rho \phi^{(n)}\left(\zeta \right)\Big) \nn 
%----------
[1.5mm] 
&&
+\,
L^{-6n(D-\Delta+1)} \,
{\cal F}^{(n)}_{0,\nu\rho\tau}
\Big( {\cal P}^{(n)}_{\mu\zeta} \dl_\sigma \phi^{(n)}\left(\zeta \right)+ {\cal P}^{(n)}_{\mu\sigma} \dl_\zeta \phi^{(n)}\left(\zeta \right)\Big) \nn
%----------
[1.5mm]
&&
+\,
L^{-6n(D-\Delta+1)} \,
\Big( {\cal F}^{(n)}_{0,\mu\zeta\sigma} {\cal Q}^{(n)}_\nu + {\cal F}^{(n)}_{0,\nu\zeta\sigma} {\cal Q}^{(n)}_\mu \Big)
\dl_\rho \phi^{(n)}\left(\zeta \right)\dl_\tau \phi^{(n)}\left(\zeta \right)\nn
%----------
[1.5mm]
&& 
+\, 
L^{-6n(D-\Delta+1)} \,
\Big( {\cal P}^{(n)}_{\mu\zeta} \dl_\sigma \phi^{(n)}\left(\zeta \right)+ {\cal P}^{(n)}_{\mu\sigma}  \dl_\zeta \phi^{(n)}\left(\zeta \right)\Big)
\Big( {\cal P}^{(n)}_{\nu\rho} \dl_\tau \phi^{(n)}\left(\zeta \right)   + {\cal P}^{(n)}_{\nu\tau}    \dl_\rho \phi^{(n)}\left(\zeta \right) \Big) \nn
%----------
[1.5mm]
&&
+\, 
L^{-6n(D-\Delta+1)} \,
 {\cal Q}^{(n)}_\nu \dl_\rho \phi^{(n)}\left(\zeta \right)\dl_\tau \phi^{(n)}\left(\zeta \right)
\Big( {\cal P}^{(n)}_{\mu\zeta} \dl_\sigma \phi^{(n)}\left(\zeta \right)+ {\cal P}^{(n)}_{\mu\sigma}  \dl_\zeta \phi^{(n)}\left(\zeta \right)\Big) \nn
%----------
[1.5mm]
&&
+\, 
L^{-6n(D-\Delta+1)} \,
{\cal Q}^{(n)}_\mu \dl_\zeta \phi^{(n)}\left(\zeta \right)\dl_\sigma \phi^{(n)}\left(\zeta \right)
\Big( {\cal P}^{(n)}_{\nu\rho} \dl_\tau \phi^{(n)}\left(\zeta \right)   + {\cal P}^{(n)}_{\nu\tau}  \dl_\rho \phi^{(n)}\left(\zeta \right) \Big) \nn
%----------
[1.5mm]
&& 
+\, 
L^{-6n(D-\Delta+1)} \,
{\cal Q}^{(n)}_\mu {\cal Q}^{(n)}_\nu 
\dl_\rho \phi^{(n)}\left(\zeta \right) \dl_\tau \phi^{(n)} \left(\zeta \right) 
\dl_\zeta \phi^{(n)}\left(\zeta \right)\dl_\sigma \phi^{(n)}\left(\zeta \right)\nn
%----------
[1.5mm]
&&
+\, 
L^{-6n(D-\Delta+1)} \,
{\cal F}^{(n)}_{0,\mu\zeta\sigma} {\cal F}^{(n)}_{0,\nu\rho\tau}
\bigg\}
\nn
%----------
[1.5mm]
&&
+\,\sigma_0^2 \, 
L^{-n(4D-4\Delta+2)} \,
\mathfrak{g}^{(n)}{}^{\sigma\tau} \mathfrak{g}^{(n)}{}^{\rho\zeta}  
\left(\dl_{\rho} \dl_{\sigma} \dl_{\tau} \phi^{(n)}\left(\zeta \right)\right)
\bigg\{
\nn
%-----
[1.5mm]
&&
\hspace{4mm}
\Big(
2 \dlx \bF^{(n)}_\mu(x)\big|_{x=\bx} \dlx \bF^{(n)}_\nu(x)\big|_{x=\bx}
+3 \sigma_0^2 \dlx^2 \bF^{(n)}_\mu(x)\big|_{x=\bx} \dlx^2 \bF^{(n)}_\nu(x)\big|_{x=\bx}
\Big)
\dl_\zeta \phi^{(n)}\left(\zeta \right)
\nn
%-----
[1.5mm] 
&&
+ \, {\cal F}^{(n)}_{1,\mu\nu\zeta}  
\bigg\}.
\end{eqnarray} 
%****************************************** 
\allowdisplaybreaks[0]
%******************************************
In the one above, $\dl_\mu=\dl/\dl \zeta^\mu$.

In $S^{(n)}$ above, we can see there appear three kinds of the exponents, 
which we express as $\kappa_{1,2}$ as
\begin{eqnarray}
\label{kdvpvdcac}
        \kappa_1 \!\! &=& \!\! {D}/{2}+4-2(D-\Delta+2), \\ 
[1.5mm] \kappa_2 \!\! &=& \!\! {D}/{2}+6-6(D-\Delta+1), \\
[1.5mm] \kappa_3 \!\! &=& \!\! {D}/{2}-(4D-4\Delta+2),
\end{eqnarray}
where in the value of $\kappa_1$, 
we have take into account of the two facts:  
1) $L^n \, \mathfrak{g}^{(n)}_{\delta\epsilon}(\zeta) \, \mathfrak{g}^{(n)}{}^{\epsilon\zeta}(\zeta)=\delta^\delta_\zeta$,  
2) We later take the contraction as mentioned under (\ref{njomzkmtr}). 
As a result not $6$ but $4$ has been taken.  
Then, if we take $\Delta$ such that 
\begin{itemize}
\item $\kappa_1$ vanishes; $\Delta=3D/4$, which leads $\kappa_2=-D$ and $\kappa_3=-2-D/2$.
\item $\kappa_2$ vanishes, $\Delta=11D/12$, which leads $\kappa_1=D/3$ and $\kappa_3=-2+D/6$.
\item $\kappa_3$ vanishes, $\Delta=(4+7D)/8$, which leads $\kappa_1=(4+D)/4$ and $\kappa_2=3-D/4$. 
\end{itemize}
Therefore, when we take $\Delta$ as
\begin{eqnarray}\label{delfxac}
\Delta = 3D/4, \quad \textrm{where $D=2$ in this study.} 
\end{eqnarray}
the fixed-point exists, which is 
\begin{eqnarray}\label{njomzkmtrac}
\lim_{n \to \infty} S^{(n)}(\zeta)    
\!\!&=&\!\!
-D 
\int d^D\zeta \sqrt{-\mathfrak{g}^{(\infty)}\left(\zeta \right)} \, 
\mathfrak{g}^{(\infty)}{}^{\mu\nu}\left(\zeta \right) \, 
\dl_\mu \phi^{(\infty)} \left(\zeta \right) \, \dl_\nu \phi^{(\infty)} \left(\zeta \right),  
\end{eqnarray}  
where we have performed the contraction: 
$
L^{2n}
\mathfrak{g}^{(n)}{}^{\sigma\tau}\left(\zeta \right)
\mathfrak{g}^{(n)}{}^{\rho\zeta}\left(\zeta \right)
\mathfrak{g}^{(n)}_{\zeta\sigma}\left(\zeta \right)  
\mathfrak{g}^{(n)}_{\rho\tau}\left(\zeta \right) = D
$. 

However, if it comes to $\phi^{(n)}(\zeta)$ with $\Delta$ in (\ref{delfxac}), 
the relations of (\ref{DefFishmet1}) and (\ref{metexpfm}) 
for the coarse-grained $\phi^{(n)}(\zeta)$ and $g^{(n)}{}^{\mu\nu}(\zeta)$ 
and $ g^{(n)}_{\mu\nu}(\zeta)$ cannot be held.  
(Only when $\Delta$ is given as $D$ as in (\ref{delfx}), it can be held.)  

\end{document}